\title{DES Science Portal: Creating Science-ready Catalogs}

\documentclass[final,5p,times,twocolumn]{elsarticle}
\usepackage{hyperref}
\usepackage{lineno}
\usepackage{ulem}
\modulolinenumbers[2]
\journal{Journal of \LaTeX\ Templates}
\usepackage{comment}
\usepackage{colortbl}
\usepackage{color}
\usepackage{aas_macros}
\usepackage{tabularx}
\usepackage{multirow}
\usepackage[colorinlistoftodos]{todonotes}

\biboptions{authoryear}

\usepackage{multicol}        % Multi-column entries in tables
\usepackage[T1]{fontenc}
\usepackage{ae,aecompl}
\usepackage{newtxtext,newtxmath}
\usepackage{listings}

\usepackage{eso-pic}% http://ctan.org/pkg/eso-pic

\AddToShipoutPictureBG*{%
  \AtPageUpperLeft{%
    \hspace{0.75\paperwidth}%
    \raisebox{-3.5\baselineskip}{%
      \makebox[0pt][l]{\textnormal{DES 2016-0176}}
}}}%

\AddToShipoutPictureBG*{%
  \AtPageUpperLeft{%
    \hspace{0.75\paperwidth}%
    \raisebox{-4.5\baselineskip}{%
      \makebox[0pt][l]{\textnormal{FERMILAB-PUB-17-088}}
}}}%

\begin{document}
    \begin{frontmatter}

    \title{DES Science Portal: Creating Science-Ready Catalogs}

\author[linea,lsst]{Angelo~Fausti~Neto\corref{cor1}}
    \ead{angelofausti@linea.gov.br}
\author[linea,on]{Luiz~N.~da~Costa\corref{cor1}}
    \ead{ldacosta@linea.gov.br}
\author[linea,on]{Aurelio~Carnero} 
\author[linea,on]{Julia~Gschwend} 
\author[linea,on]{Ricardo~L.C.~Ogando}
\author[linea,unicamp]{Flavia~Sobreira}
\author[linea,on]{Marcio~A.G.~Maia}
\author[linea,urgs]{Basilio~X.~Santiago}
\author[linea,unesp]{Rogerio~Rosenfeld}
\author[linea]{Cristiano~Singulani}
\author[linea]{Carlos~Adean}
\author[linea]{Lucas~D.P.~Nunes}
\author[linea,cefet]{Riccardo~Campisano}
\author[linea]{Rafael~Brito}
\author[linea]{Guilherme~Soares}
\author[linea]{Glauber~C.~Vila-Verde}
\author[ctio]{Tim~M.C.~Abbott}
\author[ucl,rhodes]{Filipe~B.~Abdalla}
\author[fermilab]{Sahar~Allam}
\author[ucl,cnrs,iap]{Aur\'elien~Benoit-L{\'e}vy}
\author[ucl]{David~Brooks}
\author[fermilab]{Elizabeth~Buckley-Geer}
\author[portsmouth]{Diego~Capozzi}
\author[uni-illinois,ncsa]{Matias~Carrasco~Kind}
\author[ifae]{Jorge~Carretero}
\author[upenn]{Chris~B.~D'Andrea}
\author[kandi]{Shantanu~Desai}
\author[ucl]{Peter~Doel}
\author[fermilab]{Alex~Drlica-Wagner}
\author[uni-michigan-astro,uni-michigan-phys]{August~E.~Evrard}
\author[ieec-csic]{Pablo~Fosalba}
\author[ift-uam]{Juan~Garc\'ia-Bellido}
\author[uni-michigan-astro,uni-michigan-phys]{David~W.~Gerdes}
\author[uni-illinois,ncsa]{Robert~A.~Gruendl}
\author[fermilab]{Gaston~Gutierrez}
\author[ohio-ccapp,ohio-phys]{Klaus~Honscheid}
\author[uw]{David~J.~James}
\author[usc-phys,usc-ipp]{Tesla~E.~Jeltema}
\author[aao]{Kyler~Kuehn}
\author[argone]{Steve~Kuhlmann}
\author[fermilab]{Nikolay~Kuropatkin}
\author[ucl]{Ofer~Lahav}
\author[ifusp,linea]{Marcos~Lima}
\author[tam]{Jennifer~L.~Marshall}
\author[princeton]{Peter~Melchior}
\author[uni-illinois,ncsa]{Felipe~Menanteau}
\author[jpl]{Andr\'es~Plazas}
\author[ciemat]{Eusebio~Sanchez}
\author[fermilab]{Vic~Scarpine}
\author[slac]{Rafe~Schindler}
\author[uni-michigan-phys]{Michael~Schubnell}
\author[ciemat]{Ignacio~Sevilla-Noarbe}
\author[southampton]{Mathew~Smith}
\author[ctio]{Robert~C.~Smith}
\author[ornl]{Eric~Suchyta}
\author[ncsa]{Molly~E.C.~Swanson}
\author[uni-michigan-phys]{Gregory~Tarle}
\author[ctio]{Alistair~R.~Walker}

\cortext[cor1]{Corresponding authors}
\address[linea] {Laborat\'orio Interinstitucional de e-Astronomia - LIneA, Rua General Jos\'e Cristino, 77, Rio de Janeiro, RJ, 20921-400, Brazil}
\address[lsst]{LSST Project Management Office, Tucson, AZ, USA}
\address[on] {Observat\'orio Nacional, Rua General Jos\'e Cristino, 77, Rio de Janeiro, RJ, 20921-400, Brazil} %
\address[unicamp]{Instituto de F\'isica Gleb Wataghin, Universidade Estadual de Campinas, Campinas, SP, 13083-859, Brazil}
\address[urgs]{Instituto de F\'\i sica, Universidade Federal do Rio Grande do  Sul, Caixa Postal 15051, Porto Alegre, RS - 91501-970, Brazil}
\address[unesp] {IFT-UNESP \& ICTP-SAIFR, S\~ao Paulo, SP - 01140-070, Brazil}
\address[cefet] {Centro Federal de Educa\c c\~ ao Tecnol\'ogica Celso Suckow da Fonseca - CEFET/RJ, Av. Maracan\~a, 229, Rio de Janeiro, RJ, 20271-110, Brazil}
\address[ctio]{Cerro Tololo Inter-American Observatory, National Optical Astronomy Observatory, Casilla 603, La Serena, Chile}
\address[ucl]{Department of Physics \& Astronomy, University College London, Gower Street, London, WC1E 6BT, UK}
\address[rhodes]{Department of Physics and Electronics, Rhodes University, PO Box 94, Grahamstown, 6140, South Africa}
\address[fermilab]{Fermi National Accelerator Laboratory, P. O. Box 500, Batavia, IL 60510, USA}
\address[cnrs]{CNRS, UMR 7095, Institut d'Astrophysique de Paris, F-75014, Paris, France}
\address[iap]{Sorbonne Universit\'es, UPMC Univ Paris 06, UMR 7095, Institut d'Astrophysique de Paris, F-75014, Paris, France}
\address[portsmouth]{Institute of Cosmology \& Gravitation, University of Portsmouth, Portsmouth, PO1 3FX, UK}
\address[uni-illinois]{Department of Astronomy, University of Illinois, 1002 W. Green Street, Urbana, IL 61801, USA}
\address[ncsa]{National Center for Supercomputing Applications, 1205 West Clark St., Urbana, IL 61801, USA}
\address[ifae]{Institut de F\'{\i}sica d'Altes Energies (IFAE), The Barcelona Institute of Science and Technology, Campus UAB, 08193 Bellaterra (Barcelona) Spain}
\address[upenn]{Department of Physics and Astronomy, University of Pennsylvania, Philadelphia, PA 19104, USA}
\address[kandi]{Department of Physics, IIT Hyderabad, Kandi, Telangana 502285, India}
\address[uni-michigan-astro]{Department of Astronomy, University of Michigan, Ann Arbor, MI 48109, USA}
\address[uni-michigan-phys]{Department of Physics, University of Michigan, Ann Arbor, MI 48109, USA}
\address[ieec-csic]{Institut de Ci\`encies de l'Espai, IEEC-CSIC, Campus UAB, Carrer de Can Magrans, s/n,  08193 Bellaterra, Barcelona, Spain}
\address[ift-uam]{Instituto de Fisica Teorica UAM/CSIC, Universidad Autonoma de Madrid, 28049 Madrid, Spain}
\address[ohio-ccapp]{Center for Cosmology and Astro-Particle Physics, The Ohio State University, Columbus, OH 43210, USA}
\address[ohio-phys]{Department of Physics, The Ohio State University, Columbus, OH 43210, USA}
\address[uw]{Astronomy Department, University of Washington, Box 351580, Seattle, WA 98195, USA}
\address[usc-phys]{Department of Physics, University of California,1156 High St. Santa Cruz, CA, 95064, USA}
\address[usc-ipp]{Santa Cruz Institute for Particle Physics, University of California, 1156 High St. Santa Cruz, CA, 95064, USA
}
\address[aao]{Australian Astronomical Observatory, North Ryde, NSW 2113, Australia}
\address[argone]{Argonne National Laboratory, 9700 South Cass Avenue, Lemont, IL 60439, USA}
\address[ifusp] {Departamento de F\'isica Matem\'atica, Instituto de F\'isica, Universidade de S\~ao Paulo, CP 66318, S\~ao Paulo, SP, 05314-970, Brazil }
\address[tam] {George P. and Cynthia Woods Mitchell Institute for Fundamental Physics and Astronomy, and Department of Physics and Astronomy, Texas A\&M University, College Station, TX 77843, USA}
\address[princeton]{Department of Astrophysical Sciences, Princeton University, Peyton Hall, Princeton, NJ 08544, USA}
\address[jpl] {Jet Propulsion Laboratory, California Institute of Technology, 4800 Oak Grove Dr., Pasadena, CA 91109, USA}
\address[ciemat]{Centro de Investigaciones Energ\'eticas, Medioambientales y Tecnol\'ogicas (CIEMAT), Madrid, Spain}
\address[slac]{SLAC National Accelerator Laboratory, Menlo Park, CA 94025, USA}
\address[southampton] {School of Physics and Astronomy, University of Southampton, Southampton, SO17 1BJ, UK}
\address[ornl] {Computer Science and Mathematics Division, Oak Ridge National Laboratory, Oak Ridge, TN 37831, USA}

%
%%%%%% remove this block when the affiliations problem is solved %%%%%%%%%%%    
    \begin{keyword}
{astronomical databases: catalogs, surveys} --
{methods: data analysis}
    \end{keyword}
    \end{frontmatter}

%

%
%\begin{abstract}
%
\newpage

\section*{Abstract}
We present a novel approach for creating science-ready catalogs through a software infrastructure developed for the Dark Energy Survey (DES). We integrate the data products released by the DES Data Management and additional products created by the DES collaboration in an environment known as DES Science Portal. Each step involved in the creation of a science-ready catalog is recorded in a relational database and can be recovered at any time. We describe how the DES Science Portal automates the creation and characterization of lightweight catalogs for DES Year 1 Annual Release, and show its flexibility in creating multiple catalogs with different inputs and configurations. Finally, we discuss the advantages of this infrastructure for large surveys such as DES and the Large Synoptic Survey Telescope. The capability of creating science-ready catalogs efficiently and with full control of the inputs and configurations used is an important asset for supporting science analysis using data from large astronomical surveys.
%
%   \end{abstract}
%    
%    \begin{keyword}
%{astronomical databases: catalogs, surveys} --
%{methods: data analysis}
%    \end{keyword}
%    
%    \end{frontmatter}

%\linenumbers

%%%%%%%%%%%%%%%%%%%%%%%
    \section{Introduction}
    \label{sec:intro}
%%%%%%%%%%%%%%%%%%%%%%%

Over the last decade, large and multi-wavelength photometric surveys have taken center stage as one of the main research tools in Astronomy. The need for ever increasing volumes and homogeneous statistical samples for cosmological studies, the discovery of rare populations and time-domain studies, combined with the coming of age of large and efficient mosaic cameras, have motivated a number of optical and infrared surveys. Examples in the modern era include the Sloan Digital Sky Survey \citep[SDSS,][]{Yor00}, the Two Micron All Sky Survey \citep[2MASS,][]{Skr06}, the Canada-France-Hawaii Legacy Survey \citep[CFHTLS,][]{LeF05}, the Cosmological Evolution Survey \citep[COSMOS,][]{Sco07}, the VISTA Hemisphere Survey \citep[VHS,][]{McM13}, the Kilo-Degree Survey \citep[KIDS,][]{deJ13}, the Panoramic Survey Telescope \& Rapid Response System \citep[PANSTARRS,][]{Kai10,Res14,Sco14}, the Dark Energy Survey \citep[DES,][]{Fla05}, and will culminate with the 10-year survey to be conducted with the Large Synoptic Survey Telescope \citep[LSST,][]{Ive08,Abe09}.

These surveys have had a profound impact on astronomy turning it from a data starved to a data-intensive science and forcing new methods in computer science to be developed to handle the large data volumes and complex procedures involved in preparing the data for scientific analysis. For example, SDSS is one of the most used and cited surveys in history in part due to data access interfaces like Sky Server \citep{szalay2002sdss} and CASJobs \citep{li2008casjobs}, which provide access to the SDSS data releases to the public.

Other domains such as material science, plant biology, and genomics have been more active recently in developing portals, also known as science gateways, to their communities in support of reproducibility and open access \citep{Mar13, Ges16}. For Astronomy, a few science-as-a-service pilots are emerging, such as the container-based analysis platform SciServer \citep{Rad17} and the Theoretical Astrophysical Observatory \citep{Ber16} focused on synthetic galaxy catalog production.

The DES collaboration is a 5-year program to carry out two distinct surveys. The wide-angle survey covers 5,000 deg$^2$ of the southern sky in the ($grizY$) filters to a nominal magnitude limit of $\sim$24 in most bands. Also, there is a deep survey ($i \sim$26) of about 30 deg$^2$ in four filters ($griz$) with a well-defined cadence to search for type-Ia Supernovae {(SNe Ia)} \citep{Kes15}. The primary goal of the DES is to constrain the nature of dark energy through the combination of four observational probes, namely baryon acoustic oscillations, counts of galaxy clusters, weak gravitational lensing, and determination of distances of {SNe}. Once the data are collected, the DES Data Management (DESDM) system at the National Center for Supercomputing Applications 
 \citep[NCSA \footnote{\url{http://www.ncsa.illinois.edu/}}, e.g.,][]{Des12,Moh12,Mor18}  processes the images, and produces a catalog of objects with a large number of measurements and associated masks.  The subsequent analyses rarely use all of the measurements. It usually defines new masks and ancillary data products, and sometimes apply new calibrations to the "raw" data to produce the refined catalogs that serve as input to the calculation of the science-relevant statistics.

In this paper, we address the issue of creating "science-ready" catalogs for DES Year 1 Annual Release in a manner that is traceable and reproducible given the many choices that go into producing them and the continuous evolution of versions of the data products involved.  We describe the software infrastructure developed for this purpose which is part of the DES Science Portal  (hereafter referred as "the portal", da Costa et al. 2017, in preparation) a facility, complementary to the DESDM system, meant to support scientific analysis and enhance the usability of the DES data products.

In Section~\ref{sec:overview} we present an overview of our approach to create science-ready catalogs. In Section~\ref{sec:input_data_produts} we describe the input data products such as co-addded products, ancillary maps and value-added products and how they are used. In  Section~\ref{sec:lightweight_lss} we describe how the portal automates the creation and characterization of lightweight catalogs, describing in detail the example of preparing a catalog for the study of Large Scale Structure. In Sections~\ref{sec:other_lightweight_catalogs}  to \ref{sec:special_samples} we illustrate how the infrastructure that we developed can be used to create different types of catalogs.  In Section~\ref{sec:operational_benefits} we discuss the operational benefits of this infrastructure and in Section~\ref{sec:future_developments} the future developments are mentioned. Finally, in Section~\ref{sec:summary} we summarize our results.

%%%%%%%%%%%%%
    \section{Overview}
    \label{sec:overview}
%%%%%%%%%%%%%

 The science-ready catalogs are created as part of a larger workflow in the portal. In a typical use, a scientist might:
    \begin{itemize}
\item log into the portal web interface;
\item select an input catalog from the available data releases;
\item specify in a web interface the criteria for an object to be included in the science-ready catalog;
\item specify value-added information to be included for each object;
\item execute a science-analysis pipeline using the science-ready catalog as input;
\item use additional tools available in the portal for data mining or download the results. 
\end{itemize}

The science-ready catalogs were primarily designed to feed the science-analysis pipelines in the portal. They have a reduced number of rows (objects) and columns (properties) compared to the objects catalog released by DESDM. 
    \begin{figure}
    \begin{center}
    \includegraphics[width=0.475\textwidth]{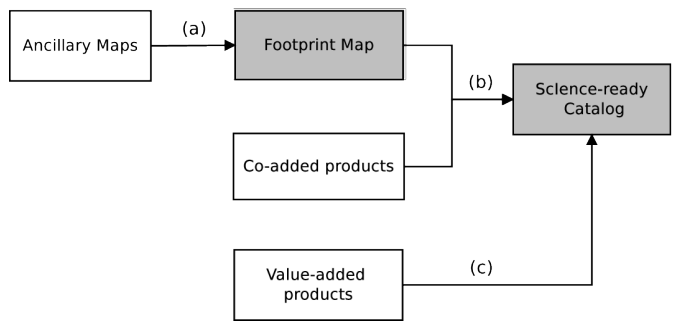}
    \caption{Simplified view of the process of creating a science-ready catalog showing the input data products (open rectangles), and the output data products (filled rectangles). The main steps involved are (a) \textit{region selection}, (b) \textit{object selection} and \textit{column selection}, and (c) the addition of value-added quantities to the final catalog. }
     \label{fig:simplified-view}
    \end{center}
    \end{figure}

 Figure~\ref{fig:simplified-view} illustrates the process of creating a science-ready catalog. In addition to the co-added products released by DESDM (the objects catalog and the \texttt{mangle} mask), other data products like ancillary maps and value-added products are required. The ancillary maps are used in the \textit{region selection} step and the result of the combination of those maps is the footprint map associated with the final catalog. When combined with the co-added objects table, the footprint map removes objects in regions affected by artifacts (such as bright stars, foreground galaxies, and globular clusters), and set the catalog area based on constraints applied on depth and other survey parameters. Constraints on the object sample like magnitude cuts, signal-to-noise cuts, color cuts, and quality flags are  applied in the \textit{object selection} step. Finally, only the relevant columns for a particular analysis are selected from the objects table. Other properties like photometric redshifts (photo-$z$s), as well as parameters from the ancillary maps, can be added to the final catalog. The result of this approach is an efficient tool that can be used by the scientist to automate the creation of science-ready catalogs, and test the impact of the different inputs and configurations on the science results. 

In the catalog infrastructure, a relational database stores an inventory of the input data products, and all steps above are implemented through SQL queries. A large number of tables and configuration parameters often result in complex SQL queries and optimization problems. This issue motivated the development of a module called {\verb query_builder }. The {\verb query_builder } creates the SQL queries for the scientist through a graphical user interface that simplifies the selection of the input data and the catalog configuration. Once a catalog is created, the selected input data, the configuration and the SQL queries executed are registered in the database and associated with the process that created the catalog. The characterization of the final catalog is then performed by another module, {\verb catalog_properties }, which creates plots of the projected distribution, number counts, color-color, color-magnitude, star--galaxy separation and photo-$z$ distributions depending on its application.

There are specific pipelines implemented in the portal to create science-ready catalogs, they share the same infrastructure but the input data and the configuration are different in each case. The science-ready catalogs are divided into three categories: i) lightweight catalogs designed to feed the science analysis pipelines (see Sections ~\ref{sec:lightweight_lss} and ~\ref{sec:other_lightweight_catalogs}) ii) generic catalogs that can be created with the same infrastructure but used for analysis outside the portal (see Section ~\ref{sec:generic_catalog}), and iii) special samples which are derived from i) with further selection criteria (see Section ~\ref{sec:special_samples}).

The reader should note that the results and statistics presented in this paper are meant to illustrate the portal infrastructure. While the services are provided to all DES collaborators, they are not necessarily the source of the science-ready catalogs used for every DES publication.

%%%%%%%%%%
    \section{Data products}
    \label{sec:input_data_produts}
%%%%%%%%%%%

Currently, the portal is running at the Laborat\'orio Nacional de Computa\c c\~ao Cient\'ifica (LNCC\footnote{\url{http://www.lncc.br/}}). As part of the DES public data release (DR1) effort, we plan to migrate some of the portal services to NCSA. In the meantime, the co-added products released by DESDM must be transferred to LNCC and uploaded into the portal. This task is done in an early stage called \textsc{Data Installation}. 

For the DES Year 1 Annual Release (hereafter, Y1A1), some data products derived from the single epoch data or the co-added products, such as the ancillary maps and value-added products, were created by the portal during an intermediate stage called \textsc{Data Preparation}, while others were prepared by the DES collaboration and uploaded. Ideally, all data products would be created through the portal for rapid turn around when new data releases are available. Our expectation is to create the data products increasingly through the portal, as the survey science collaboration matures.

In the portal, the co-added products are organized by Data Release (in this case Y1A1) and Data Set. Data Set corresponds to independent fields in this release such as the wide fields overlapping the SDSS Stripe 82 (S82) region, the South Pole Telescope region \citep[SPT,][]{Car11}, and the supplemental fields at different depths.

%%%%%%%%%%%%%
    \subsection{Co-added products}
    \label{sec:coadded_products}
%%%%%%%%%%%%%

DESDM provides the co-added objects catalog and the footprint masks in the \texttt{mangle} \citep{Swa12} format. The Y1A1 co-added objects table contains 139,142,161 unique objects spread over 1,800 deg$^2$ in two wide regions S82 and SPT. The Y1A1 co-adds are combinations of up to 5 exposures in each of the $grizY$ filters. The typical coverage across the footprint is about $N=3.5$ exposures in each filter.

Y1A1 also contains supplemental fields. Many individual exposures of the DES Supernovae fields (C, S, X, E), the COSMOS \citep{Sco07} fields, and the VVDS14 \citep{LeF05} fields were obtained during year-1 observations and DES Science Verification (SV). These exposures have been co-added into three sets of $90$  tiles\footnote{A DES tile is a region of coadded data on sky spanning an area of 0.5 deg$^ 2$.} in three separate depths: D04 (similar to the depth of the Y1A1 S82 and SPT regions), D10 (equivalent to $N=10$ representing a 5-year completed survey depth) and DFULL (with all exposures available from SV and year-1). In each set, there are 74 SN tiles, 8 COSMOS tiles, and 8 VVDS14 tiles. The supplemental fields are used for the creation of training sets for photo-$z$ algorithms \citep{Gsc17}. 

The Y1A1 \texttt{mangle} masks have about $10^8$ distinct polygons representing the detailed geometry of the observations and the co-addition, keeping information about co-added weight, effective area, magnitude limit and exposure time of the survey as a function of the position on the sky. Another mask, the bit mask, is used to eliminate objects that overlap bright stars and bleed trails, also represented as \texttt{mangle} masks.

%%%%%%%%%%%%%%
    \subsection{Ancillary maps}
    \label{sec:ancillary_maps}
%%%%%%%%%%%%%%

In addition to the co-added products, ancillary maps were produced to characterize the coverage, depth and observing conditions of the survey. An important result of the process of creating a science-ready catalog is the footprint map which is created by combining several ancillary maps.

The Y1A1 ancillary maps are Hierarchical Equal Area isoLatitude Pixelation \citep[\texttt{HEALPix},][]{Go05}  \footnote{\url{http://healpix.sourceforge.net/index.php}} representations of the survey characteristics as described by \cite{Drl18}. By default, we use \texttt{NSIDE}=4096 corresponding to a pixel area of 0.73 arcmin$^2$. \texttt{HEALPix} supports two different numbering schemes for the pixels, \texttt{NESTED} and \texttt{RING}; in the current implementation only \texttt{RING} is used. While the \texttt{HEALPix} maps are not exact representations of the survey characteristics they are computationally fast and convenient for creating the footprint map, in particular when several maps with different constraints are used (see Section ~\ref{sec:region_selection}). 

We can summarize the ancillary maps used during the creation of a science-ready catalog as:

\begin{itemize}

\item Coverage fraction map (hereafter,  \texttt{detfrac}
maps) are used in the \textit{region selection} step to remove non-observed regions and to compute the catalog area. They are created at the working resolution of \texttt{NSIDE}=4096 by computing the fraction of subpixels at higher resolution (\texttt{NSIDE}=32768, pixel area of 0.01 arcmin$^2$) that are contained within the \texttt{mangle} mask \citep{Drl18}. For example, a pixel with \texttt{DETFRAC\_I}=0.8 at \texttt{NSIDE}=4096 means that 80\% of its area has been observed.

\item Bad region mask is designed to remove catalog artifacts like unphysical colors, astrometric discrepancies, bright stars, large foreground galaxies and bright galaxies. See Table~\ref{table:flag_codes} for a list and description of the flags used. For the complete description of each flag and the criteria used to define the area removed by the various flags see \cite{Drl18}. When the bad region mask was first used in association with the Cluster Finder pipeline in the portal, we noticed a significant number of spurious galaxy cluster detections due to globular clusters. Thus we added a flag based on the globular cluster catalog of \citet{Har10} to remove those regions. In the future, we propose to separate the current bad region mask in two, one to flag artifacts associated with the release and another one to flag regions affected by Foreground Objects, so that these products can be created independently and the Foreground Objects mask reused in subsequent data releases.

\item Survey depth maps give the magnitude limit at 5-$\sigma$ and 10-$\sigma$ as a function of position on the sky for the \texttt{AUTO} and \texttt{APER4} magnitudes \citep{Ryk15}.

\item Systematics (or survey conditions) maps contain the total exposure time, mean seeing, mean air mass and sky background in each of the $grizY$ filters. Their creation is implemented in the portal following the prescription of \citet{Lei16}. 

\end{itemize}

    \begin{table*}%[!b]
     \footnotesize
    \begin{center}
    \renewcommand{\arraystretch}{1.25}
    \caption{Bad region mask flags.}
    \label{table:flag_codes}
    \vspace{0.25cm}
    \begin{tabular}{cl}
    \hline
    \hline
    \textbf{Flag \#} & \textbf{Description}  \\  
    \hline
    \ 1 & High density of astrometric discrepancies \\
    \ 2 & 2MASS moderate star regions ($8<J<12$) \\
    \ 4 & RC3 large galaxy region ($10<B<16$) \\
    \ 8 & 2MASS bright star region ($5<J<8$) \\
    16 & Region near the LMC \\
    32 & Yale bright star region ($-2<V<5.6$) \\
    64 & High density of unphysical colors \\
    128 & Globular cluster regions from \citet{Har10} catalog \\
    \hline
    \end{tabular}
    \end{center}
    \end{table*}
%

%%%%%%%%%%%%%%%%%%
    \subsection{Value-added products}
    \label{sec:value-added_products}
%%%%%%%%%%%%%%%%%%

The value-added products used as input to create catalogs are the zeropoint correction, photo-$z$s, star-galaxy separation and galaxy evolution properties computed in the \textsc{Data Preparation} stage for each object in the co-added objects table.
Table~\ref{table:properties} summarizes the data products and the properties stored in the database for each value-added product.

%%%%%%%%%%%%%%%%%%%%%
    \subsubsection{Zeropoint correction}
    \label{sec:zeropoint}
%%%%%%%%%%%%%%%%%%%%%    

  The Y1A1 data is calibrated using a Global Calibration Module (GCM) developed by DES\footnote{\url{https://github.com/DarkEnergySurvey/GCM}}, which follows the procedure of \citep{Gla94} adapted to DES by \citet{Tuc07}. 
However, internal studies have shown that Y1A1 residual calibrations  uncertainties at the level of 2\% persist. A stellar locus regression (SLR) solution has been applied to the data for zeropoint calibration, following the prescription detailed in \citet{Ive04} and \citet{Mac04}.

In the portal, the zeropoint correction includes the correction of the magnitudes by extinction \citep[SFD98][]{Sch98}, and finding an SLR solution at the scale of a DES tile ($0.5$ deg$^2$) using a modified version of the \texttt{BigMACS}\footnote{\url{https://code.google.com/archive/p/big-macs-calibrate/}} SLR code \citep{Kel14}.

Differences in calibration might affect color cuts, photo-$z$ estimations, and the star-galaxy separation. Therefore, the same calibration must be applied consistently to the different data products that are used as input to create a science-ready catalog. Photometric consistency is ensured by the provenance scheme built into the portal. For future DESDM data releases, different calibration techniques might be applied reinforcing the importance of keeping track of this information.

%%%%%%%%%%%
    \subsubsection{Photo-zs}
    \label{sec:photoz}
%%%%%%%%%%%
    
The pipelines implemented in the portal to compute the photo-$z$s are fully described by \citet{Gsc17}. The steps involved are summarized below:

\begin{itemize}

\item creation of a spectroscopic database, currently with redshift measurements for a total of 31 galaxy spectroscopic surveys available in the literature, and 759,890 unique, high-quality spectroscopic redshifts;

\item creation of a spectroscopic sample by combining the surveys, homogenizing the data and resolving multiple redshift measurements of the same source; 

\item matching of the spectroscopic sample with the co-added objects table used as input to create training and validation sets; 

\item training of the photo-$z$ algorithms; 

\item calculation of the photo$z$s with the three algorithms used in this paper: \texttt{DNF} \citep{deV15}, \texttt{LePhare} \citep{Arn02} and \texttt{MLZ/TPZ} \citep{Car13,Car14}.  

\end{itemize}

For interoperability among the different algorithms, the original property name in the output of each algorithm is translated to the portal's internal format when the database table is created. The name of the algorithm used is registered along with other metadata associated with the process.
 
Even though some photo-$z$ algorithms compute Probability Density Functions (PDFs) it can be computationally very expensive to store them for all the objects in the co-added objects table \citep[e.g.,][]{Car14}. Within the portal, one way to alleviate this is to compute and store photo-$z$ PDFs for the science-ready catalogs or \textit{Special Samples} (see Section~\ref{sec:special_samples}) which are reduced in size compared to the whole sample. 

    \begin{table*}
    \footnotesize
    \begin{center}
    \renewcommand{\arraystretch}{1.25}
    \caption{Description of the value-added products and their properties stored in the database in the \textsc{Data Preparation} stage.} 
    \label{table:properties}
    \vspace{0.25cm}
    \begin{tabular}{ll}
    \hline
    \hline
    \textbf{Zeropoint correction} & \\
    \hline
    \texttt{COADD\_OBJECTS\_ID} & Unique object identifier\\
    \texttt{SLR\_SHIFT} & SLR magnitude shifts for each of the $grizY$ filters \\
    \texttt{EXTINCTION} & Extinction for each of the $grizY$ filters \\
    \hline
    \textbf{Photo-$z$s} & \\
    \hline
    \texttt{COADD\_OBJECTS\_ID} & Unique object identifier\\
    \texttt{Z\_BEST} & Best estimate of the photo-$z$ \\
    \texttt{ERR\_Z} & Photo-$z$ error \\  
    \hline
    \textbf{Star-galaxy separation} &  \\
    \hline
    \texttt{COADD\_OBJECTS\_ID} & Unique object identifier\\
    \texttt{CLASS\_STAR} & Star-galaxy classification\textsuperscript{$\dagger$} ($0=$ galaxy and $1=$ star) for each of the $grizY$ filters \\
    \texttt{SPREAD\_MODEL} & Star-galaxy classification{$\ddagger$}  \\
    \texttt{MODEST} & Star-galaxy classification\textsuperscript{$\diamondsuit$}, v1 and v2 which are based on the \texttt{SPREAD\_MODEL} \\
    \hline
    \textbf{Galaxy evolution properties} &  \\
    \hline
    \texttt{COADD\_OBJECTS\_ID} & Unique object identifier\\
    \texttt{Z\_BEST} & Best estimate of the photo-$z$ \\
    \texttt{MAG\_ABS} & Absolute magnitude \\
    \texttt{K\_COR}  &  k-correction  \\
    \texttt{DIST\_MOD\_BEST}  &  Distance modulus  \\
    \texttt{MASS\_BEST}  & Stellar mass for the best galaxy model  \\
    \texttt{SFR\_BEST}  &  Star-formation rate for the best galaxy model \\
    \texttt{AGE\_BEST}  &  Age of stellar population for the best galaxy model \\
    \texttt{EBV\_BEST}  &  Internal extinction for each of the $grizY$ filters  \\
    \hline
    \multicolumn{2}{l}{{$\dagger$}\footnotesize{\ \citet{Ber96} }} \\
    \multicolumn{2}{l}{{$\ddagger$}\footnotesize{\ \citet{Des12}}} \\
    \multicolumn{2}{l}{{$\diamondsuit$}\footnotesize{\ \citet{Cha15}}} \\
    \end{tabular}
    \end{center}
    \end{table*}
%

%%%%%%%%%%%%%%
    \subsubsection{Star-galaxy separation}
    \label{sec:sg_sep}
%%%%%%%%%%%%%%
   
The star-galaxy separation algorithms developed by the collaboration are described by \citet{Sev18}.
Five algorithms are currently implemented in the portal:  \texttt{CLASS\_STAR} \citep{Ber96}, \texttt{SPREAD\_MODEL} \citep{Des12} and Y1A1 \texttt{MODEST} v1 and v2 \citep[e.g.,][]{Cha15} which are also based on the \texttt{SPREAD\_MODEL} parameter. So far, those star-galaxy separation algorithms use only morphological information. Thus the classification is indeed between extended and point sources, referred here as galaxies and stars respectively. However, the infrastructure can be extended to include other classes of objects, e.g., QSO, as other classification algorithms are implemented. As in the case of other value-added products, the output of each algorithm must be translated to a common format within the portal to ensure interoperability.

%%%%%%%%%%%%%%%%%%%
    \subsubsection{Galaxy evolution properties}
    \label{sec:galprops}
%%%%%%%%%%%%%%%%%%%

For galaxy evolution studies, galaxy properties are computed by the \texttt{LePhare} \citep{Arn02} algorithm at the redshift \texttt{Z\_BEST} computed either by \texttt{LePhare} itself or by another photo-$z$ algorithm implemented in the portal. The galaxy evolution properties for the best magnitude model solution based on the Spectral Energy Distribution (SED) used as template are listed in Table~\ref{table:properties} as well.

%%%%%%%%%%%%%%%
    \section{Use case example: lightweight catalog for Large Scale Structure}
    \label{sec:lightweight_lss}
%%%%%%%%%%%%%%%    

To illustrate the creation of a science-ready catalog in the portal, we use as an example a lightweight magnitude-limited catalog adequate for computing the angular correlation function. The portal graphical user interface helps the user select the input data and configuration (see ~\ref{app:user_interfaces}). For the \textit{Large Scale Structure} (LSS) pipeline, the data products presented as input to the user are the co-added objects table, the star-galaxy separation and the photo-$z$ tables for the different algorithms implemented in the portal. By selecting those products, the methods for star-galaxy separation and photo-$z$ are immediately set. Once the input data are selected, the user has the chance to change the default configuration parameters for the LSS catalog before submitting a process in the portal. At this point, the input data and the configuration used are saved and associated with the new process.

Next, we describe in detail how the {\verb query_builder } performs the \textit{region selection}, \textit{object selection} and \textit{column selection} steps to create the LSS catalog (see Figure ~\ref{fig:query_builder}). In the  ~\ref{app:query_builder} we present the  SQL queries created for this particular example.
%%%%%%%%%%%%
    \subsection{Region selection}
    \label{sec:region_selection}
%%%%%%%%%%%%

We start by creating a footprint map as the result of the constraints applied to the \texttt{detfrac} maps, bad region mask, systematics maps and depth maps.  

For the LSS default configuration (see  Table~\ref{tab:region_selection}), only pixels with \texttt{DETFRAC\_I} $>0.8$ are selected.  The bad region mask with \texttt{FLAG=2,4,8,32,128} ensure that regions affected by bright stars, large foreground galaxies, and globular clusters are removed from the catalog.  From the systematics maps, we use the exposure time map to select pixels with \texttt{EXPTIME} $>=90$s in each of the $griz$ filters  \footnote{The exposure time is $90$s in the $griz$ filters and $45$s in the $Y$ filter.}. In this example, we keep only pixels in the depth map with magnitude limit $i>22$, so that our final catalog can include galaxies with signal-to-noise of 10 $\sigma$ or higher at magnitudes brighter than $i=22$ \footnote{The limiting magnitude used in this example is conservative and is just to illustrate the infrastructure.}.  The appropriate depth map is chosen from the selected resolution (\texttt{NSIDE}=4096), magnitude-type (\texttt{MAG\_AUTO}) and signal-to-noise of the limiting magnitude (10$\sigma$). The current infrastructure relies on the depth maps created by the collaboration that includes SLR adjustments and extinction correction  (see details in \cite{Drl18}). Therefore, in order to be consistent, the magnitudes used in the preparation of training sets, in computing photo-$z$s and listed in the final catalog must be corrected accordingly. This need shows the importance of having the depth maps also created through the portal for self-consistency. 

%%%%%%%%%%%%%%%
    \subsection{Object selection}
    \label{sec:object_selection}
%%%%%%%%%%%%%%%

% the fraction 16% is found by dividing the area of the footprint map by the area of the original co-added objects table, e.g., for S82 140.6 sq deg and 167 sq deg respectively and taking the complementary number.

The resulting footprint map is then combined with the co-added objects table, which removes about $16\%$ of the objects in this operation. For the LSS default configuration (see Table~\ref{tab:object_selection}), the selected sample includes objects with magnitudes $17.5<i<22$; colors $-1<g$-$r<3$, $-1<r$-$i<2.5$, $-1<i$-$z<2$, $-5<z$-$Y<5$; and \texttt{SExtractor} quality \texttt{FLAG} $=0$, $1$, and $2$ in the $i$-filter. As discussed in the \textit{region selection} step, in this example only pixels with \texttt{DETFRAC\_I} $>0.8$ are kept. Then the \texttt{mangle} mask in the $i$-filter must be applied to make sure that the objects in the selected pixels that overlap the \texttt{mangle} mask are properly removed. 

Still in the \textit{object selection} step, there are additional cuts that are applied by default to remove artifacts associated with stars close to the saturation threshold, objects with bad astrometric colors and objects with unphysical colors (see details in \cite{Drl18}).
 
The resulting object sample is then combined with the Y1 \texttt{MODEST} v2 star-galaxy separation table to select objects classified as galaxies using the value of the classifier in the $i$-filter as a reference. As in the \textit{region selection} step, all these parameters are presented in the configuration interface and can be changed by the user (see Figure~\ref{fig:configuration}).

%%%%%%%%%%%%%%%
    \subsection{Column selection}
    \label{sec:column_selection}
%%%%%%%%%%%%%%%

For the LSS lightweight catalog, a few columns are selected by default. These are called system default since they are required to feed the science analysis pipelines that use this catalog. The columns are: \texttt{COADD\_OBJECTS\_ID}, \texttt{RA}, \texttt{DEC}, \texttt{MAG\_[GRIZY]}, \texttt{MAGERR\_[GRIZY]}, \texttt{Z\_BEST} and \texttt{ERR\_Z} with magnitudes and errors consistent with the magnitude type selected in the \textit{object selection} configuration. Still in the \textit{column selection} step, additional columns from the co-added objects table, value-added products or ancillary maps can be selected in the configuration interface and added to the final catalog.

The execution time for the \textit{region selection}, \textit{object selection} and \textit{column selection} steps is roughly proportional to the number of objects in the co-added objects table. In this example, the execution time for the S82 region (334 tiles) was 00h22m while for SPT (3,373 tiles) it was 02h45m. It is important to note that the overall execution time to create a science-ready catalog was significantly reduced also because the value-added products were previously computed for all the objects in the co-added objects table. In that process, the photo-z estimation is the most time consuming step \citep[see][]{Gsc17}.

    \begin{table}
    \footnotesize
    \caption{Properties of the S82 and SPT magnitude-limited catalogs.}
    \label{tab:maglim_properties}
    \begin{center}
    \renewcommand{\arraystretch}{1.25}
    \begin{tabular}{lccc}
    \hline
    \hline
    \textbf{Catalog} & \textbf{Footprint Area } & \textbf{Ngals} &     \textbf{Mean Density } \\
    & (deg$^2$) & & (gal arcmin$^{-2}$) \\
    \hline
S82 & 140.65  & 1,806,274 & 3.57 \\
SPT & 1,375.48 & 17,915,328 & 3.63 \\
    \hline
    \end{tabular}
    \end{center}
    \end{table}
%

% Process 717 was used for S82  and process 720 used for SPT
\begin{figure*}[htpb]
    \centering
    \includegraphics[width=0.98\textwidth]{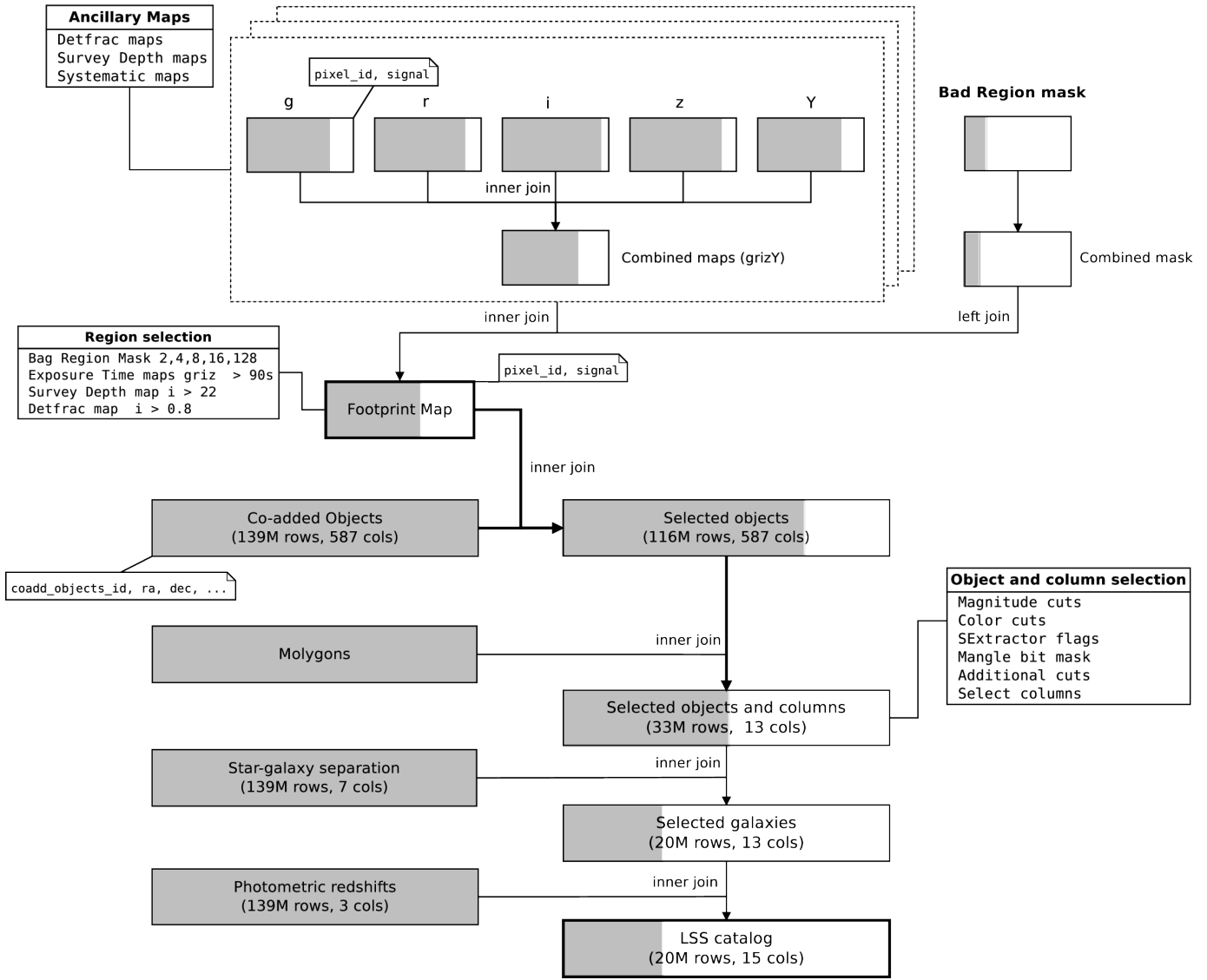}
    \caption{Representation of the SQL operations executed during the creation of the LSS lightweight catalog described in Section~\ref{sec:lightweight_lss}. Rectangles are input data or temporary tables created along the process. The ancillary maps on the top are combined to create maps with constraints on the  \texttt{detfrac}, survey depth and systematics maps parameters. The combined maps are then joined with the bad region mask to create the footprint map. The shaded area represents qualitatively how the size of the catalog is shrinking (number of rows and columns) in each step. The join of the footprint map with the co-added objects table removes a large number of objects, speeding up the subsequent \textit{object selection} step. Joins with the value-added products and the co-added objects sample are done only at the end, operating on a reduced object sample for efficiency. The resulting catalog is optimized for its science application.
    }
    \label{fig:query_builder}
    \end{figure*}
%

%%%%%%%%%%%%%%
    \subsection{Characterization of the catalog properties}
    \label{sec:properties}
%%%%%%%%%%%%%%
 
After creating a catalog, the portal performs an automatic characterization of its properties. In this section, we illustrate some of the results of this characterization applied to our LSS lightweight catalog.

\begin{figure*}
    \begin{center}
    \includegraphics[angle=0,width=1.2\columnwidth]{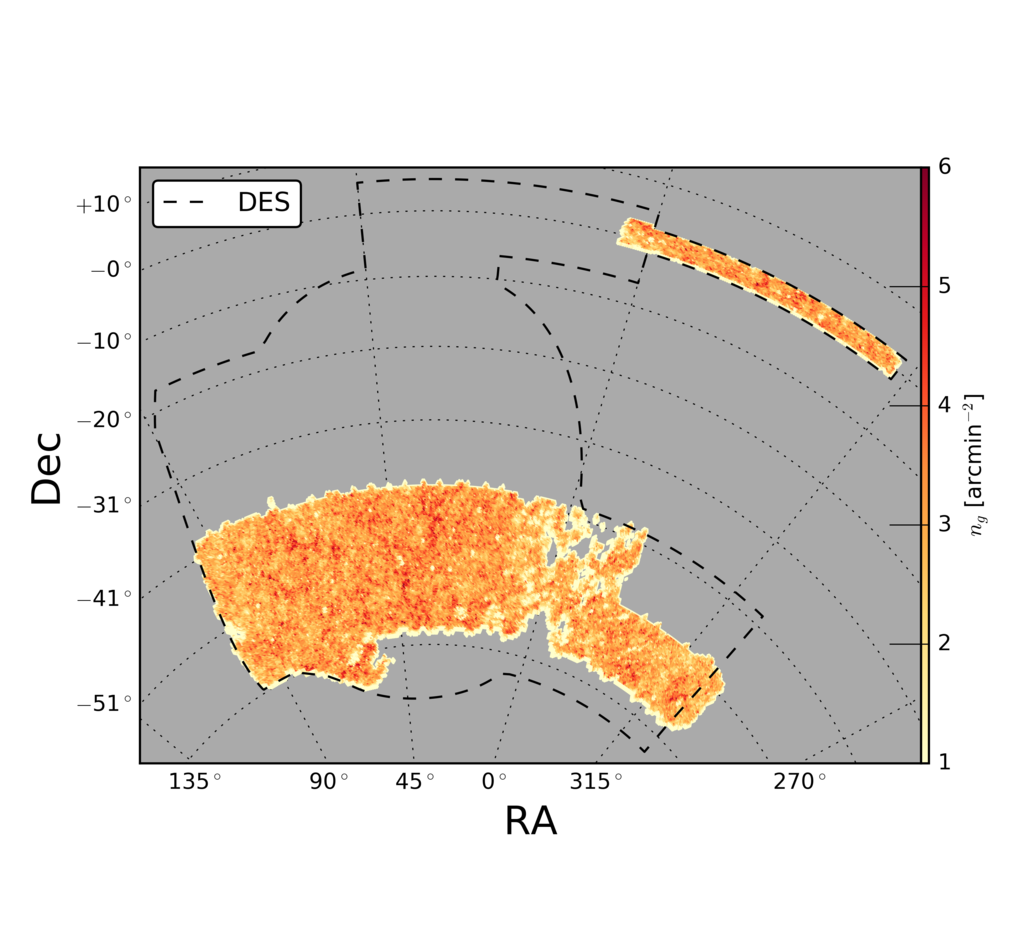}   
    \end{center}
    \caption{Projected density distribution for the S82 (around Dec$=0^{\circ}$) and SPT (-60$^{\circ}<$ Dec $<-40^{\circ}$) regions of the lightweight LSS catalog.}  
    \label{fig:maglim_projected}
    \end{figure*}

The basic properties of our LSS lightweight catalog are listed in Table~\ref{tab:maglim_properties} which gives the area of the final footprint, the number of galaxies and the mean density of galaxies per arcmin$^2$ in the final catalog. The projected density distribution is shown in Figure~\ref{fig:maglim_projected}. Close inspection of the table and figure indicates that the catalogs are pretty uniform across the sky, and have similar mean number densities, differing by less than 1\%.

\begin{figure}[h]
    \begin{center}
    \begin{minipage}{0.45\textwidth}
    \includegraphics[width=\textwidth]{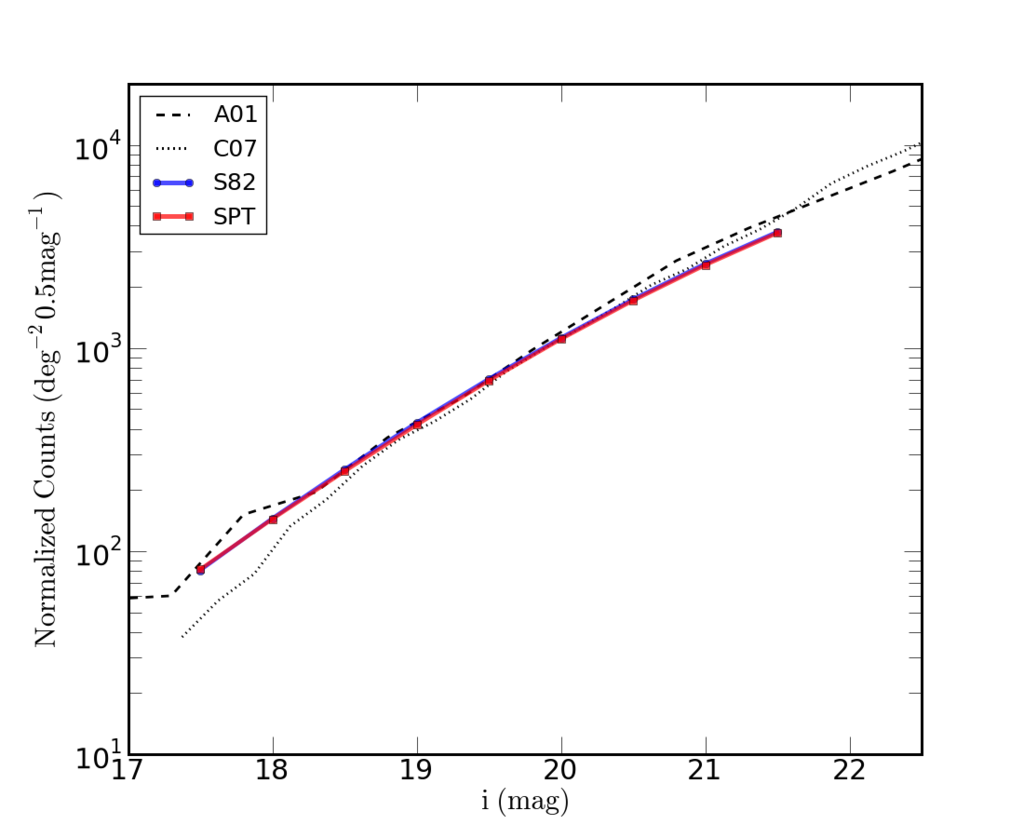}
    \end{minipage}
    \end{center}
    \caption{Normalized counts of galaxies in the $i$-band for the S82 and SPT magnitude-limited catalogs. We also show results obtained by other authors: A01- \citet{Arn01}, and C07- \citet{Cap07}.    }
    \label{fig:maglim_nc}
    \end{figure}

In Figure~\ref{fig:maglim_nc} we compare the $i$-band magnitude distributions for both S82 and SPT regions with those obtained by other authors, normalized to the same area. We find excellent overall agreement between the counts of the two independent regions and consistency with the counts of the other authors despite possible differences in the $i$-filter used.

    \begin{figure}[h]
    \begin{center}  
    \begin{minipage}[b]{0.45\textwidth}
    \includegraphics[width=0.84\textwidth]{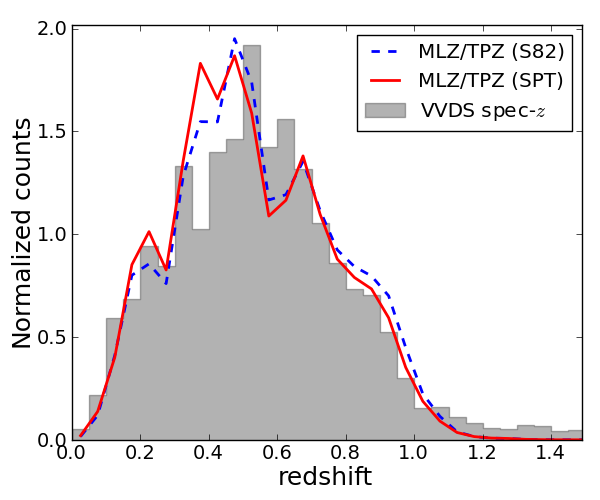} 
    \end{minipage}
    \end{center}
    \caption{Distribution of photo-$z$s computed by the \texttt{MLZ/TPZ} algorithm for the S82 (blue line) and SPT (red line) magnitude-limited catalogs. The gray histogram represents the distribution of VVDS spectroscopic survey for the same magnitude limit ($i=22.0$).}
    \label{fig:maglim_photoz}
    \end{figure}

The photo-$z$ distribution computed by the \texttt{MLZ/TPZ} algorithm for the S82 and SPT regions is shown in Figure~\ref{fig:maglim_photoz}. From the figure we see that the photometric distributions of the two regions are similar and in reasonable agreement with that of VVDS spectroscopic survey limited to the same magnitude, except by the excess in the 0.3-0.5 interval. 

We conclude that the automatic characterization built on the portal is useful to quickly assess the self-consistency of catalogs created with the same configuration but using different input data. In this case, the disjoint S82 and SPT regions in Y1A1 demonstrates the overall uniformity of the survey, and  the good agreement with the results of other surveys.

The automatic characterization, as well as the record of the input data and configurations, are especially important when multiple catalogs are created, as discussed in Section~\ref{sec:impact}. 

%%%%%%%%%%%
     \subsection{Creating multiple catalogs}
    \label{sec:impact}
%%%%%%%%%%%

Our LSS lightweight catalog was created using specific star-galaxy separation and photo-$z$ algorithms. In order to explore the possible impact of these choices on the science results, we demonstrate the value of the portal in creating multiple catalogs for the S82 region by changing the methods used for separating stars and galaxies and for computing photo$z$s.

\begin{figure}
    \begin{center}
    \begin{minipage}[b]{0.5\textwidth}
    \includegraphics[width=\textwidth]{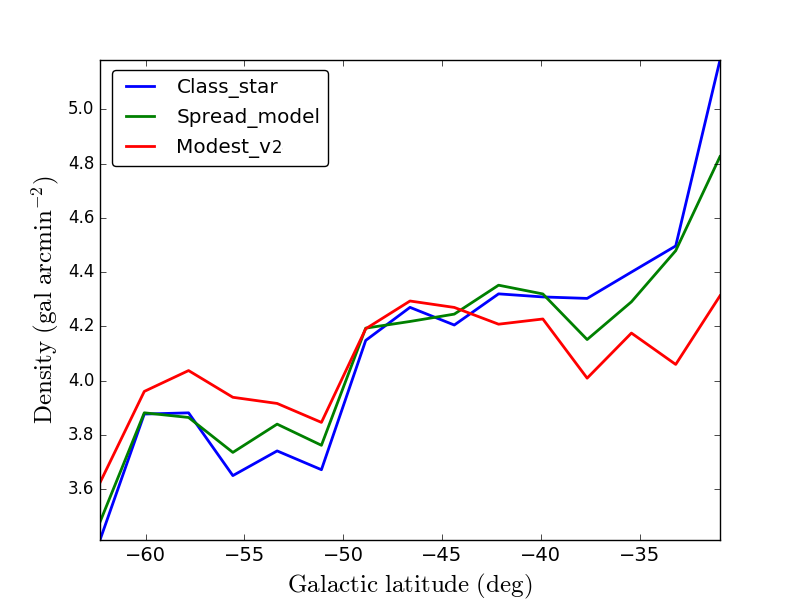}
    \end{minipage}
    \end{center}
    \caption{Impact of different star-galaxy separation algorithms in the density of galaxies as a function of the galactic latitude, $b^{\textsc{ii}}$.}
     \label{fig:sg_impact_s82}
    \end{figure}

The effects of using different methods for separating stars and galaxies can be examined in Figure~\ref{fig:sg_impact_s82}, which shows the variation of the projected density of galaxies as a function of the galactic latitude, $b^{\textsc{ii}}$. The methods used were \texttt{CLASS\_STAR}, \texttt{SPREAD\_MODEL} and Y1 \texttt{MODEST} v2. They show a rapid rise in the density for galactic latitudes below $b^{\textsc{ii}}\sim-35^{\circ}$, suggesting some degree of contamination by stars wrongly classified as galaxies.  Although it is not a quantitative way to establish the best star-galaxy separation algorithm, it at least shows that our choice was reasonable given the alternatives.

\begin{figure}
    \begin{center}
    \begin{minipage}[b]{0.5\textwidth}
    \includegraphics[width=0.84\textwidth]{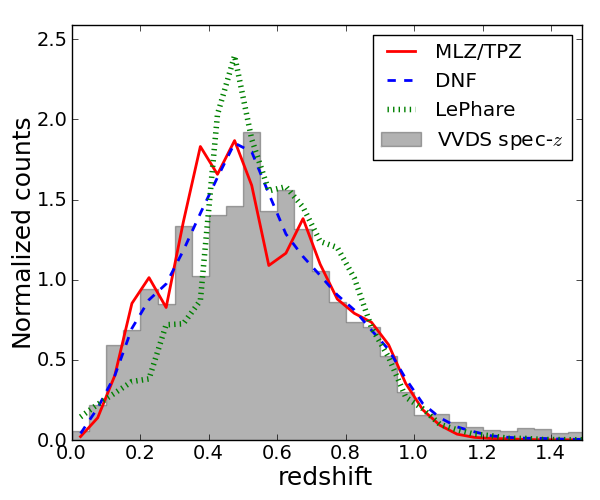}
    \end{minipage}
    \end{center}
    \caption{Distribution of the photo-$z$s computed by the \texttt{MLZ/TPZ} (red), \texttt{DNF} (blue) and \texttt{LePhare} (green) algorithms for the catalog described in Section~\ref{sec:lightweight_lss} for the SPT region.}  
    \label{fig:photoz_impact_spt}
    \end{figure}

Similarly, to assess the impact of using different photo-$z$ algorithms we show in Figure~\ref{fig:photoz_impact_spt} the redshift distribution for three catalogs created using \texttt{MLZ/TPZ}, \texttt{DNF} and \texttt{LePhare} for the SPT region.  From the figure, we find that the empirical methods yield similar results over the entire range. They contrast with the SED fitting method which deviates considerably from the empirical methods for $z<0.5$. Interestingly, all methods yield very similar distributions for $z>0.7$.

The important point is that the present infrastructure allows the user to generate different catalogs quickly, feed the science analysis pipelines implemented in the portal with those catalogs and evaluate the impact that different inputs and configurations have on the scientific results. Something like that would be costly to do by hand especially with the increasing volume of data. In addition, the same infrastructure could be easily adapted to create catalogs from simulated data in order to assess the performance of the star galaxy-separation and photo-z algorithms against a truth table.

%

% From processes 510, 511 and 525
%
%

%%%%%%%%%%%%%%
    \section{Other lightweight catalogs}
    \label{sec:other_lightweight_catalogs}
%%%%%%%%%%%%%%

In section~\ref{sec:lightweight_lss} we described how we use the portal to create a lightweight catalog adequate for LSS studies with just the columns required to compute the angular correlation function in the portal. The {\verb query_builder } is flexible enough to create catalogs for different applications, changing only the input data products and the configuration used. As in the case of LSS, catalogs for \textit{Galaxy Clusters} (Cluster), \textit{Galaxy Evolution} (GE) and \textit{Galaxy Archaeology} (GA)  are also lightweight and designed to feed the corresponding analysis pipelines in the portal ready to be used and with only the required columns. The default configuration for the lightweight catalogs is summarized in ~\ref{app:query_builder}. As explained in Section~\ref{sec:operational_benefits}, the user can change and save specific configurations for each pipeline using the Configuration Manager user interface.

For cluster studies, we created catalogs (similar to the LSS ones) to feed the \textit{Cluster Finder} pipeline implemented in the portal. This pipeline uses the Wavelet Z Photometric \citep[\texttt{WAZP}][]{Ben18} cluster-finding algorithm, which is based on the 3D spatial clustering considering both the projected distribution of galaxies and their photo-$z$ distribution.  The default Cluster catalog has the same columns as the LSS one, and the only differences in the configuration are in the bright magnitude and color cuts, as shown in ~\ref{app:query_builder}.

% process id 10025381

For galaxy evolution studies, we have created magnitude-limited catalogs adding columns from the galaxy properties table (see Section~\ref{sec:value-added_products}) containing estimates for the stellar mass, absolute magnitude, star-formation rate, spectral type, age of stellar population, internal extinction and k-correction.  The GE lightweight catalog presented here was created for the S82 dataset using the \texttt{PEGASE2}\footnote{\url{ftp://ftp.iap.fr/pub/from\_users/pegase/PEGASE.2/}} set of spectral energy distribution (SED) to estimate galaxy stellar masses.

In this example, the photo-$z$s are from the \texttt{MLZ/TPZ} algorithm, which were used by \texttt{LePhare} to compute the galaxy  properties. The star-galaxy separation method used was Y1 \texttt{MODEST} v2. The apparent magnitude limits were  $ 17.5 < i < 22.0 $, resulting in a sample with 1,357,319 galaxies with absolute magnitudes in the range  $ -25 < M_{i} < -14 $ covering 143.76 square degrees.

\begin{figure*}
    \begin{center}
     \begin{minipage}[b]{\textwidth}
    \includegraphics[width=0.45\textwidth]{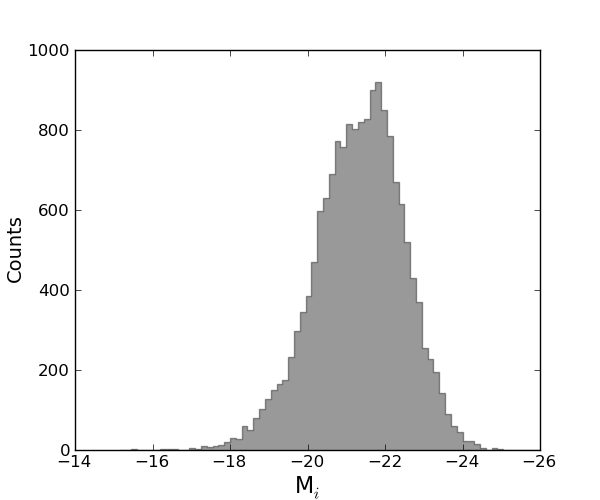}
    \includegraphics[width=0.5\textwidth]{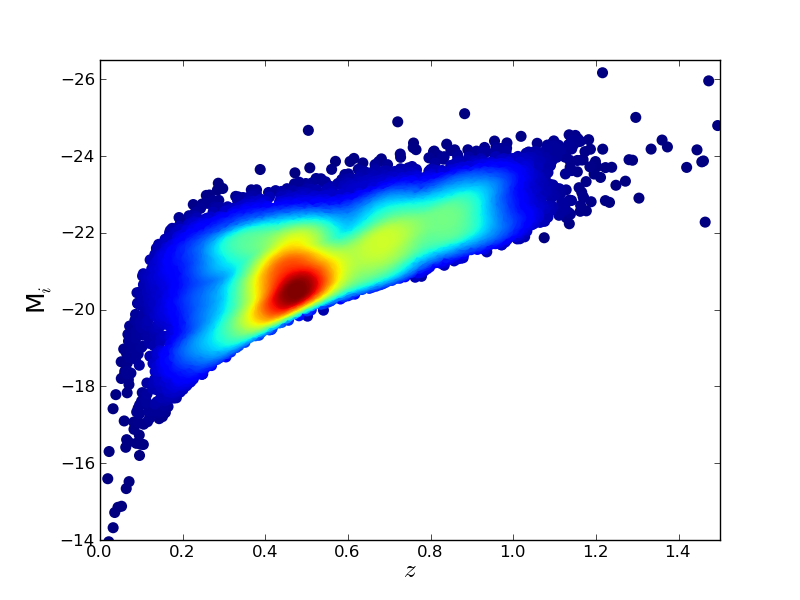}
    \end{minipage}
     \begin{minipage}[b]{\textwidth}
    \includegraphics[width=0.45\textwidth]{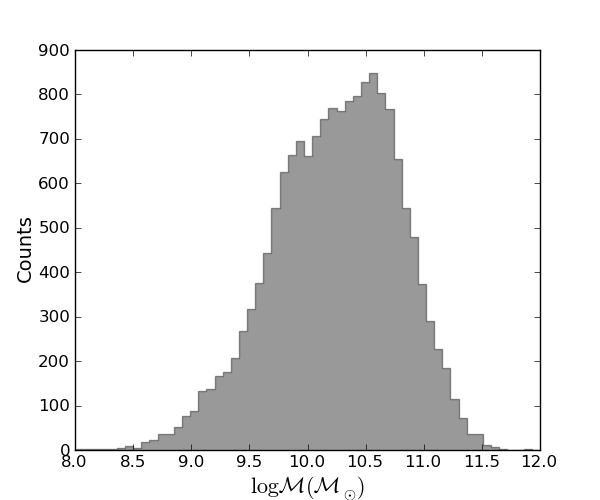}
    \includegraphics[width=0.5\textwidth]{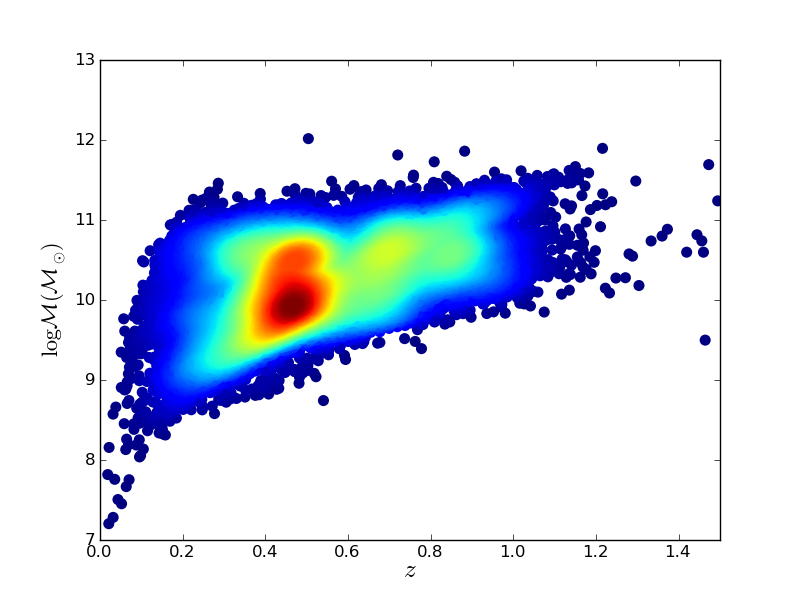}
    \end{minipage}  
    \caption{Properties of the lightweight GE catalog using default configuration as described in Section~\ref{sec:other_lightweight_catalogs}. Upper panels: distribution of absolute magnitudes and its dependence with the photo-$z$s; Lower panels: stellar mass distribution and its dependence with the photo-$z$s.  }
        \label{fig:lightweight_ge}
    \end{center}
    \end{figure*}

Figure~\ref{fig:lightweight_ge} illustrates some of the properties of our default GE catalog, showing distributions of absolute magnitudes and stellar masses and their dependence with the photo-$z$s.  The catalog contains objects from $10^{7}$ to $10^{12}$ M$_{\odot}$, with the mass distribution peaking at $10^{10.5}$ M$_{\odot}$. The ages cover the range of $\sim$100 Myr to 13 Gyr. 

In the future, we plan to include other SED libraries (e.g. \citep{Bru03} in the portal and \citet{Mar05}), in addition to PDFs associated with the photo-$z$s to better estimate their uncertainties and the impact on the luminosity and mass functions.

    \begin{figure}[h]
    \begin{center}
    \begin{minipage}[b]{0.5\textwidth}
    \includegraphics[width=\textwidth]{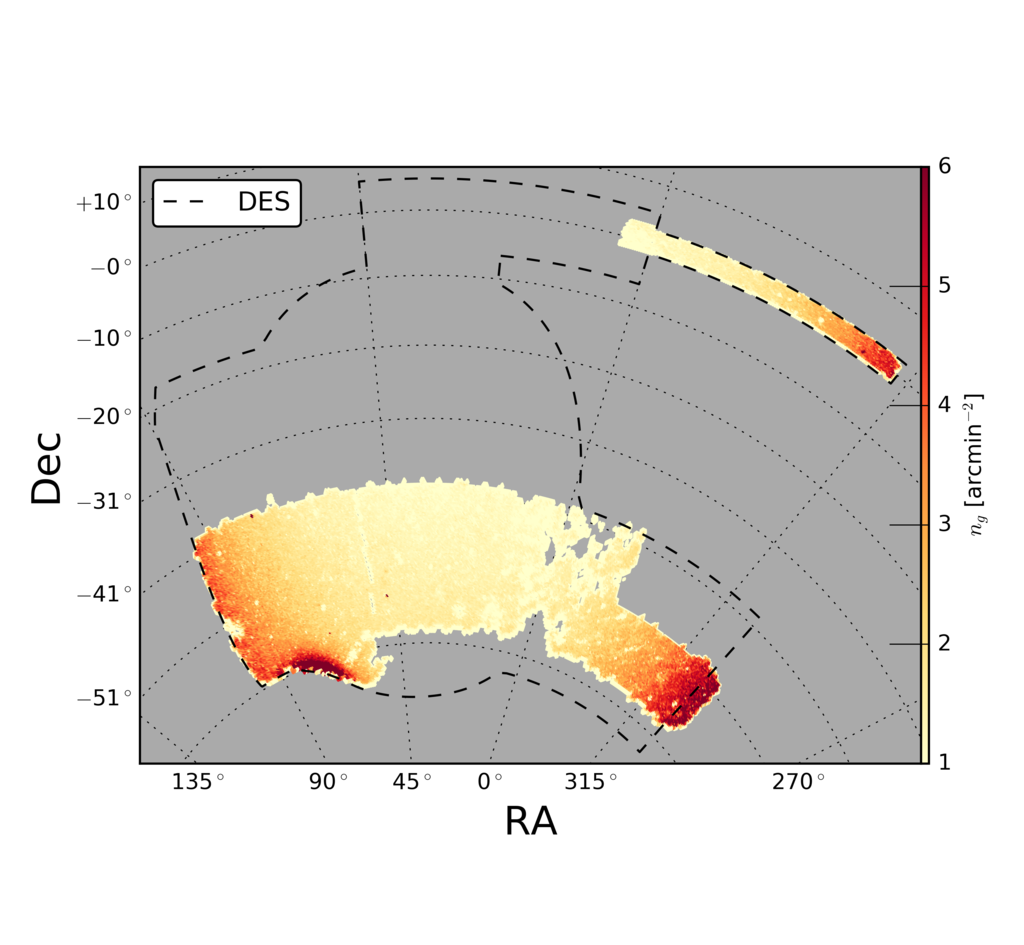}
    \end{minipage}
    \end{center}
    \caption{Density map of the lightweight GA catalog created to feed the \textit{SPARSEx} pipeline as described in Section~\ref{sec:other_lightweight_catalogs}}  
    \label{fig:lightweight_ga}
    \end{figure}

For GA studies, we created catalogs using two different configurations to feed the \textit{SPARSEx} and \textit{MWfitting} pipelines. These pipelines are also being integrated into the portal and are briefly described in the following.

The \textit{SPARSEx} pipeline is used to detect stellar systems such as globular clusters and nearby dwarf galaxies and has been successfully used with both single-epoch and co-added data \citep{Luq16,Luq17}. It applies a matched filter technique using data from the color-magnitude diagram (CMD) to build maps of stellar overdensities associated with different simple stellar population models. These overdensities are then detected in the maps and ranked according to their amplitude. Those overdensities most conspicuous and robust to variations in the model parameters are then inspected by eye and have their density profiles, and CMDs analyzed. By default the pipeline uses \texttt{WAVG\_MAG\_PSF} magnitudes for the \textit{object selection} step and the Y1 \texttt{MODEST} v2 for star-galaxy separation. The catalog to feed \textit{SPARSEx} is limited at $r=23$ at $10\sigma$. In this case, the depth map used was based on an aperture magnitude, without extinction correction. In Figure~\ref{fig:lightweight_ga} we show the footprint and the projected density distribution of this catalog.

The \textit{MWfitting} pipeline was developed to study the structure of the Galaxy. It uses models from TRIdimensional modeL of thE GALaxy \citep[\texttt{TRILEGAL},][]{Girardi2005,Girardi2012}, whose color-magnitude diagrams are computed for different structural models of the Galaxy and a best-fit solution to the observed CMDs is found based on automatic optimization algorithms. The catalog to feed this pipeline requires the selection of regions brighter than a user-specified magnitude in $r$ and $g$ filters to ensure completeness in the $g$-$r$ color. In this case, we used the depth maps in $r$ and $g$ filters, again without extinction correction. We note that like in the previous lightweight catalogs described in this section, the selected columns are just the ones required for the analysis and are listed in \ref{app:query_builder}.

%%%%%%%%%%%%%
    \section{Generic catalog}
    \label{sec:generic_catalog}
%%%%%%%%%%%%%

The available infrastructure also allows for the creation of generic catalogs suitable to be used outside the portal. For instance, these catalogs may have multiple photo-$z$s, star-galaxy classifications, and galaxy properties, leaving the decision about which method to use to the user. The \textit{region selection} and \textit{object selection} steps are still performed but neither star-galaxy classification nor photo-$z$s limits are applied. Additional columns chosen from the co-added objects table and properties associated with the various ancillary maps can be added to the catalog in this \textit{column selection} step. 

The default configuration for a generic catalog proposed in \ref{app:query_builder} is the one that creates a magnitude-limited galaxy catalog. Note that because a generic catalog might have more columns by construction and because star-galaxy separation is not applied, they are, generally, larger in size than the lightweight catalogs by a factor of two or more. However, they can still be an interesting option to enable users to carry out their science analysis outside the portal.

%%%%%%%%%%%%%%%%%%%%%%
    \section{Special samples}
    \label{sec:special_samples}
%%%%%%%%%%%%%%%%%%%%%%

In addition to the lightweight and generic catalogs described above, the portal also supports the creation of specialized samples, starting with LSS, Cluster, GE or GA catalogs as input, and applying further selections. For instance, a volume-limited catalog with information about absolute magnitude can easily be created from the GE catalog. Other use cases are the creation of Emission Line Galaxies or  Luminous Red Galaxies samples using selection criteria proposed by \citet{Com16} and \citet{LRG16}, respectively. 

As mentioned earlier, at this point one may also re-run photo-$z$ algorithms to store the PDF which is considerably more efficient given the reduced object sample. For instance, the lightweight GE catalog presented in Section~\ref{sec:other_lightweight_catalogs} has about $37\%$ of the original co-added objects used to compute point-value photo-$z$s. With this pipeline, it is also possible to combine those samples with other surveys available in the portal (such as WISE and VHS) and complement the DES data with near-infrared photometry. Again, the join operations in the database and positional matching work faster on reduced object samples.

%%%%%%%%%%%%%%% 
    \section{Operational benefits of the infrastructure}
    \label{sec:operational_benefits}
%%%%%%%%%%%%%%%

The infrastructure presented here was designed primarily to prepare catalogs to feed the science analysis pipelines hosted by the portal. The underlying idea is to make sure that i) all steps that create the input catalog are done before executing the pipelines; ii) we can control the impact of different inputs and configurations on the science results; and iii) the science analysis pipelines run on lightweight catalogs. This feature is important to reduce the data sizes and maximize performance.

As shown in Sections~\ref{sec:lightweight_lss} and \ref{sec:other_lightweight_catalogs} there are about fifty parameters defining a particular catalog and several configurations are allowed for each of the LSS, Cluster, GE, and GA pipelines. The portal has a configuration manager used to save, load and share all these different configurations. These configurations are also part of the provenance of a given portal product.

Since the execution time of some processes may be very large,  upon submission of a process the user is notified by e-mail, which contains information about the process. The same happens at the completion of the process when the user receives another notification with concise information about the configuration used, input and output data and links to pages displaying the configuration and the product log. The product log is a compilation of information about the current process and links to the previous ones that generated the input data. These links allow the user to access the whole chain of preceding processes, again including information about the inputs, configurations, version of the codes used, as well as plots and tables describing the results of the process in the chain.

Products from a given pipeline are assigned a running number which provides a unique identification for the Data Release and Data Set in addition to the name chosen by the user. Products can be published, and in this case, they automatically appear in an interface called Science Products, which users can download them (see ~\ref{app:user_interfaces}). Similarly, processes and products can be accessed from a dashboard available to the operator and system administrators but in contrast to the Science Products interface all processes are registered, even the ones that failed, thus providing a complete history of the pipeline executions. The dashboard used to monitor the execution of each pipeline in the portal is shown in Figure~\ref{fig:dashboard}. In the first tier, it shows the start time, duration and status of the latest run. In the second tier, it provides access to all previous runs with links to the product log and the data products created by each process in a third tier.

Currently, the catalog production described in this paper is only available at LNCC instance of the portal. Nevertheless, all products can be made available via a Science Products interface being developed at NCSA. This procedure is done by using the export tool that transfers any data product created by the portal to NCSA creating the corresponding table in the DES Science database.  

%%%%%%%%%%%%%%%%%
    \section{Future developments}
    \label{sec:future_developments}
%%%%%%%%%%%%%%%%%

While the current infrastructure has been extensively tested and is  already in operation, a number of improvements are necessary to keep the portal current with the algorithms and procedures defined by the DES collaboration. Some of them are:

\begin {itemize}

\item Improve the {\it Install Catalogs} pipeline not only to transfer and ingest the co-added products released by DESDM but also distribute the data among the cluster nodes partitioning the data using HEALPix.

\item Implement the local creation of depth maps. These are currently being created by the DES collaboration and uploaded to the portal. Local implementation would also allow the portal to be more flexible, enabling the creation of depth maps at different signal-to-noise ratios, with or without extinction correction and for different magnitude types.

\item Separate the current bad region mask into two, one to flag artifacts associated with the release and the other to flag regions affected by foreground objects, so that these products can be created independently and the foreground objects mask reused in subsequent releases.

\item Include other methods for computing photo-$z$ and star-galaxy separation as suggested by the DES collaboration.

\item Extend the training of photo-$z$ samples based on multi-band photometric data to complement the infrastructure based on spectroscopic samples.

\item Introduce additional SEDs to calculate galaxy properties.

\item Store $z_{MC}$, a Monte Carlo value sampled from the photo-$z$ PDF, and store a compressed representation of the PDF for each object as well.

\item Expand the number of queries available in special samples. 

\item Use of ancillary maps to create more realistic catalogs based on simulations.

\item Enable the download of data products through the Science products interface, in collaboration with the DESDM group at NCSA.

\item Ability to run the \texttt{query\_builder} in the  Jupyter\footnote{Jupyter is a web application that allows the user to create and share documents that contain live code,  visualizations, and text. \url{http://jupyter.org/}} environment. We plan to implement the configuration interface,  as well as the plots and tables for the catalog characterization in a Jupyter notebook. 
\end{itemize}

In addition to these specific short-term goals, we are currently adapting the {\verb query_builder } to work with other database engines using \texttt{SQLAlchemy}\footnote{\url{http://www.sqlalchemy.org/}}. This action is an important step to migrate the catalog infrastructure to NCSA and integrate it with the DES Science database. We are also constantly reviewing the portal code base and evaluating how to operate the portal in environments other than a dedicated cluster to avoid scalability problems in the future.

%%%%%%%%%%%
    \section{Summary}
    \label{sec:summary}
%%%%%%%%%%%

In this paper, we describe an infrastructure to create science-ready catalogs implemented in the DES Science portal. The portal creates science-ready catalogs starting from the co-added products released by DESDM, integrating algorithms and procedures developed by the DES collaboration for the Y1A1 data release. The infrastructure uses ancillary maps that describe the survey characteristics, different methods for computing star-galaxy separation, photo-$z$s and galaxy properties. 

The input data products and the configurations used are registered in a relational database. The science-ready catalogs are fully created in the database by a set of SQL queries. A module called {\verb query_builder } automatically creates the queries based on the input data products and configuration selected through the portal user interface. Provenance information is registered for each pipeline of the catalog infrastructure and is accessible through a dashboard interface making the entire process reproducible. We demonstrated the flexibility of the infrastructure by creating lightweight catalogs for LSS, Cluster, GE, and GA to feed science analysis pipelines in the portal, generic catalogs, and special samples. 

The portal makes the complex process of creating science-ready catalogs manageable, well-documented and sustainable. While this approach has been primarily motivated to feed the science analysis pipelines being implemented in the portal, the catalogs created can also be distributed for the DES collaboration. Our goal is to migrate this infrastructure to NCSA where the DES data releases are produced, turn it into an operational science environment for the DES collaboration and continue integrating new algorithms and methodologies developed or suggested by the collaboration as the survey progresses.

    \section*{Acknowledgements}

\addcontentsline{toc}{section}{Acknowledgements}

We are grateful for the extraordinary contributions of our CTIO colleagues and the DECam Construction, Commissioning and Science Verification teams in achieving the excellent instrument and telescope conditions that have made this work possible. The success of this project also relies critically on the expertise and dedication of the DES Data Management group. 

JG is supported by CAPES. ACR is supported by CNPq grant 157684/2015-6.

Funding for the DES Projects has been provided by the U.S. Department of Energy, the U.S. National Science Foundation, the Ministry of Science and Education of Spain, the Science and Technology Facilities Council of the United Kingdom, the Higher Education Funding Council for England, the National Center for Supercomputing Applications at the University of Illinois at Urbana-Champaign, the Kavli Institute of Cosmological Physics at the University of Chicago, the Center for Cosmology and Astro-Particle Physics at the Ohio State University, the Mitchell Institute for Fundamental Physics and Astronomy at Texas A\&M University, Financiadora de Estudos e Projetos, Funda{\c c}{\~a}o Carlos Chagas Filho de Amparo {\`a} Pesquisa do Estado do Rio de Janeiro, Conselho Nacional de Desenvolvimento Cient{\'i}fico e Tecnol{\'o}gico and the Minist{\'e}rio da Ci{\^e}ncia, Tecnologia e Inova{\c c}{\~a}o, the Deutsche Forschungsgemeinschaft and the Collaborating Institutions in the Dark Energy Survey. 

The Collaborating Institutions are Argonne National Laboratory, the University of California at Santa Cruz, the University of Cambridge, Centro de Investigaciones Energ{\'e}ticas, Medioambientales y Tecnol{\'o}gicas-Madrid, the University of Chicago, University College London, the DES-Brazil Consortium, the University of Edinburgh, 
the Eidgen{\"o}ssische Technische Hochschule (ETH) Z{\"u}rich, 
Fermi National Accelerator Laboratory, the University of Illinois at Urbana-Champaign, the Institut de Ci{\`e}ncies de l'Espai (IEEC/CSIC), 
the Institut de F{\'i}sica d'Altes Energies, Lawrence Berkeley National Laboratory, the Ludwig-Maximilians Universit{\"a}t M{\"u}nchen and the associated Excellence Cluster Universe, the University of Michigan, the National Optical Astronomy Observatory, the University of Nottingham, The Ohio State University, the University of Pennsylvania, the University of Portsmouth, SLAC National Accelerator Laboratory, Stanford University, the University of Sussex, Texas A\&M University, and the OzDES Membership Consortium.

Based in part on observations at Cerro Tololo Inter-American Observatory, National Optical Astronomy Observatory, which is operated by the Association of Universities for Research in Astronomy (AURA) under a cooperative agreement with the National Science Foundation.

The DES data management system is supported by the National Science Foundation under Grant Numbers AST-1138766 and AST-1536171.
The DES participants from Spanish institutions are partially supported by MINECO under grants AYA2015-71825, ESP2015-88861, FPA2015-68048, SEV-2012-0234, SEV-2016-0597, and MDM-2015-0509, some of which include ERDF funds from the European Union. IFAE is partially funded by the CERCA program of the Generalitat de Catalunya. Research leading to these results has received funding from the European Research
Council under the European Union's Seventh Framework Program (FP7/2007-2013) including ERC grant agreements 240672, 291329, and 306478. We  acknowledge support from the Australian Research Council Centre of Excellence for All-sky Astrophysics (CAASTRO), through project number CE110001020.

This manuscript has been authored by Fermi Research Alliance, LLC under Contract No. DE-AC02-07CH11359 with the U.S. Department of Energy, Office of Science, Office of High Energy Physics. The United States Government retains and the publisher, by accepting the article for publication, acknowledges that the United States Government retains a non-exclusive, paid-up, irrevocable, world-wide license to publish or reproduce the published form of this manuscript, or allow others to do so, for United States Government purposes.

This paper has gone through internal review by the DES collaboration.

    \appendix
%

%%%%%%%%%%%%%
    \section{Query builder}
    \label{app:query_builder}
%%%%%%%%%%%%%

The {\verb query_builder } was developed to automatically create the SQL queries from selected input data and configuration. The concept behind the {\verb query_builder } is that a complex query can be solved by breaking it down into simpler sub-queries creating temporary tables in the database. This procedure reduces the overall execution time ensuring that the sub-queries are written efficiently forcing the execution of the sub-queries in the right order and not depending exclusively on the database query optimizer. 

Figure~\ref{fig:query_builder} illustrates how the queries to create the LSS catalog described in Section~\ref{sec:lightweight_lss} 
are built. Because join operations on map tables are much faster than the equivalent operation on object tables, we start by operating on the ancillary maps. As described in Section~\ref{sec:region_selection}, the footprint map is what is left after the removal of pixels that satisfy the constraints imposed by the ancillary maps. The join with the footprint map removes a significant number of objects from the co-added objects table, speeding up the \textit{object selection} queries that follow. The association between \texttt{PIXEL\_ID} and \texttt{COADD\_OBJECTS\_ID} is done through an auxiliary table created using the  \texttt{PG\_HEALPIX}\footnote{\url{https://github.com/segasai/pg\_healpix}} Postgresql plugin for the appropriate resolution and ordering schema. In addition, the \textit{object selection} step is optimized by creating temporary tables with only the subset of columns required to apply the cuts described in Section~\ref{sec:object_selection}. Finally, joins involving large tables (like the co-added objects, the star-galaxy separation, and photo-$z$ tables) are expensive operations. However, performing them in individual subqueries by first removing objects through the \textit{object selection} cuts, then selecting galaxies and only then joining with the the photo-$z$ table, reduces the execution time significantly. This recipe (including the tables and parameters from the configuration, the operations and the order in which the operations are performed) is encoded in the \texttt{query\_builder}. Improvements in the \texttt{query\_builder} include expression of the operations in a configuration file to avoid changes in the code to add new operations, and rewrite the code using SQLAlchemy to build the SQL clauses in Python and support different SQL dialects like Postgresql, MySQL, and Oracle. 

    \begin{figure}
    \begin{center}
    \begin{minipage}[l]{0.485\textwidth}
    \includegraphics[width=\textwidth]{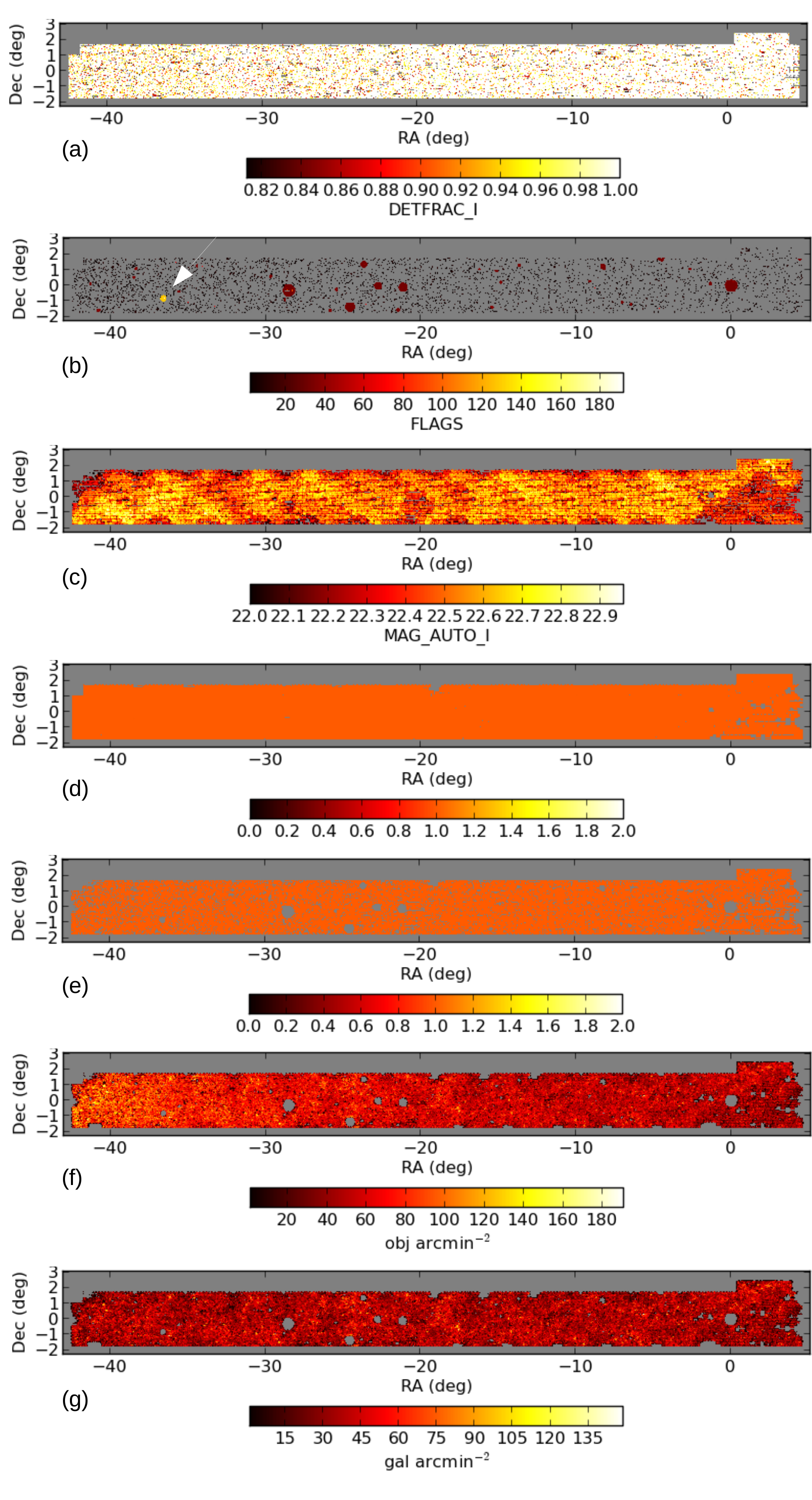}
    \end{minipage}
    \end{center}
    \caption{Steps to create the LSS catalog as described in  Section~\ref{sec:lightweight_lss} for the S82 dataset. Panel (a): regions with \texttt{DETFRAC\_I} > 0.8. Panel (b): bad region mask with \texttt{FLAG=2,4,8,32, and 128}. The white arrow indicates the position of the globular cluster M2 removed by the \texttt{FLAG=128}. Panel (c): regions with depth  with $i>22$ at 10-$\sigma$. Panel (d): binary map showing regions with \texttt{EXPTIME} $>90$s in the $griz$ filters. Panel (e): footprint map after combining the previous conditions, with a total area of 140.65 deg$^2$. Panel (f) map showing the density of objects after the \textit{object selection} step with 5.89 obj/arcmin$^2$. Panel (g): map showing the density  of galaxies after the star-galaxy separation step with 3.57 gal/arcmin$^2$. }
    \label{fig:maglim_s82}
    \end{figure}

The queries used to create the LSS catalog are presented below and can be inspected together with Figure~\ref{fig:maglim_s82}. This figure shows the ancillary maps, the resulting footprint map after the \textit{region selection} step and the density map of the selected galaxies after the \textit{object selection} step for the S82 dataset.

\lstset{
  language=sql,
  basicstyle=\footnotesize\ttfamily,
  columns=fullflexible,
  keepspaces=true,
  tabsize=4,    
}

\hspace{1cm}

1. \texttt{detfrac} map queries:
\begin{lstlisting}
CREATE TEMP TABLE <tmp_deftrac_i> AS (
    SELECT a.pixel, a.signal, a.ra, a.'dec' 
    FROM  <detfrac_i> a
    WHERE signal >= 0.8);
\end{lstlisting}

2. Bad region mask queries:
\begin{lstlisting}
CREATE TEMP TABLE <tmp_badregion> AS (
    SELECT a.pixel, a.signal, a.ra, a.'dec'
    FROM <badregion_mfask> a
    WHERE (CAST(signal AS INTEGER) & 174) > 0);
\end{lstlisting}

3. Depth map queries:
\begin{lstlisting}
CREATE TEMP TABLE <tmp_depth> AS (
    SELECT a.pixel, a.signal, a.ra, a.'dec'
    FROM <depth_i> a
    WHERE signal >= 22);
\end{lstlisting}

4. Systematic map queries:
\begin{lstlisting}
CREATE TEMP TABLE <tmp_exptime_g> AS ( 
    SELECT a.pixel, a.signal, a.ra, a.'dec' 
    FROM  <exptime_g> a
    WHERE signal >= 90);
    
CREATE TEMP TABLE <tmp_exptime_r> AS (
    SELECT a.pixel, a.signal, a.ra, a.'dec' 
    FROM  <exptime_r> a
    WHERE signal >= 90);
    
CREATE TEMP TABLE <tmp_exptime_i> AS (
    SELECT a.pixel, a.signal, a.ra, a.'dec' 
    FROM  <exptime_i> a
    WHERE signal >= 90);
    
CREATE TEMP TABLE <tmp_exptime_z> AS (
    SELECT a.pixel, a.signal, a.ra, a.'dec'
    FROM  <exptime_z> a
    WHERE signal >= 90);

CREATE TEMP TABLE <combined_exptime> AS (
    SELECT a.pixel, 1 as signal, a.ra, a.'dec'
    FROM <tmp_exptime_g> a
    INNER JOIN <tmp_exptime_r> b ON a.pixel = b.pixel
    INNER JOIN <tmp_exptime_i> c ON b.pixel = c.pixel
    INNER JOIN <tmp_exptime_z> d ON c.pixel = d.pixel);
\end{lstlisting}

5. Footprint map queries:
\begin{lstlisting}
CREATE TEMP TABLE <tmp_intersection> AS (
    SELECT a.pixel, 1 as signal, a.ra, a.'dec'
    FROM <combined_exptime> a 
    INNER JOIN <tmp_depth> b ON a.pixel = b.pixel
    INNER JOIN <tmp_detfrac> c ON b.pixel = c.pixel
);

CREATE TEMP TABLE <footprint_map> AS (
    SELECT a.pixel, a.signal, a.ra, a.'dec', b.signal
    AS detfrac_i
    FROM <tmp_pixels> a
    INNER JOIN <tmp_detfrac_i> b ON a.pixel = b.pixel
    INNER JOIN <tmp_intersection> c ON b.pixel = c.pixel
    LEFT JOIN <tmp_badregion> d ON c.pixel = d.pixel
    WHERE c.pixel IS NULL);
\end{lstlisting}

\newpage

6. Object selection queries:
\begin{lstlisting}
CREATE TEMP TABLE <tmp_reduction> AS (
    SELECT <coadd_objects>.coadd_objects_id,
           <coadd_objects>.ra,
           <coadd_objects>.dec,
           b.pixel, 
           CASE <coadd_objects>.mag_auto_g WHEN 99 THEN 99
           ELSE <coadd_objects>.mag_auto_g -
           <coadd_objects>.xcorr_sfd98_g + c.slr_shift_g 
           END AS mag_auto_g,
           CASE <coadd_objects>.mag_auto_r WHEN 99 THEN 99
           ELSE <coadd_objects>.mag_auto_r -
           <coadd_objects>.xcorr_sfd98_r + c.slr_shift_r 
           END AS mag_auto_r,
           CASE <coadd_objects>.mag_auto_i WHEN 99 THEN 99 
           ELSE <coadd_objects>.mag_auto_i -
           <coadd_objects>.xcorr_sfd98_i + c.slr_shift_i 
           END AS mag_auto_i,
           CASE <coadd_objects>.mag_auto_z WHEN 99 THEN 99
           ELSE <coadd_objects>.mag_auto_z -
           <coadd_objects>.xcorr_sfd98_z + c.slr_shift_z 
           END AS mag_auto_z,
           CASE <coadd_objects>.mag_auto_y WHEN 99 THEN 99
           ELSE <coadd_objects>.mag_auto_y -
           <coadd_objects>.xcorr_sfd98_y + c.slr_shift_y 
           END AS mag_auto_y,
           <coadd_objects>.magerr_auto_g,
           <coadd_objects>.magerr_auto_r,
           <coadd_objects>.magerr_auto_i,
           <coadd_objects>.magerr_auto_z,
           <coadd_objects>.magerr_auto_y,
           <coadd_objects>.mu_eff_model_g,
           <coadd_objects>.mu_eff_model_r,
           <coadd_objects>.mu_eff_model_i,
           <coadd_objects>.mu_eff_model_z,
           <coadd_objects>.mu_eff_model_y,
           <coadd_objects>.nepochs_g,
           <coadd_objects>.mag_model_i,
           <coadd_objects>.niter_model_g,
           <coadd_objects>.niter_model_r,
           <coadd_objects>.niter_model_i,
           <coadd_objects>.niter_model_z,
           <coadd_objects>.spreaderr_model_g,
           <coadd_objects>.spreaderr_model_r,
           <coadd_objects>.spreaderr_model_i,
           <coadd_objects>.spreaderr_model_z,
           <coadd_objects>.alphawin_j2000_i,
           <coadd_objects>.alphawin_j2000_g,
           <coadd_objects>.deltawin_j2000_g,
           <coadd_objects>.deltawin_j2000_i,
           <coadd_objects>.flags_i,
           b.pixel
    FROM <coadd_objects>
    INNER JOIN <coadd_objects_pixel> a
    ON <coadd_objects>.coadd_objects_id = a.coadd_objects_id
    INNER JOIN <footprint_map> b ON a.pixel = b.pixel
    INNER JOIN <slr> c 
    ON <coadd_objects>.coadd_objects_id =c.coadd_objects_id);
    
CREATE TEMP TABLE <tmp_cuts> AS (
    SELECT * FROM <tmp_reduction>
    WHERE mag_auto_i < 22 
    AND mag_auto_i > 17.5
    AND (((flags_i = '0') 
    OR ((CAST(flags_i AS INTEGER) & '1') > 0)
    OR ((CAST(flags_i AS INTEGER) & '2') > 0))) 
    AND mag_auto_g - mag_auto_r BETWEEN -1.0 AND 3.0 
    AND mag_auto_r - mag_auto_i BETWEEN -1.0 AND 2.5 
    AND mag_auto_i - mag_auto_z BETWEEN -1.0 AND 2.0 
    AND mag_auto_z - mag_auto_Y BETWEEN -5.0 AND 5.0 
    AND (nepochs_g > 0 or magerr_auto_g > 0.05 
    OR (mag_model_i - mag_auto_i) > -0.4) 
    AND (niter_model_g > 0 AND niter_model_r > 0 
    AND niter_model_i > 0 AND niter_model_z > 0) 
    AND (spreaderr_model_g > 0 AND spreaderr_model_r > 0 
    AND spreaderr_model_i > 0 AND spreaderr_model_z > 0) 
    AND (ABS(alphawin_j2000_g - alphawin_j2000_i) < 0.0003 
    AND ABS(deltawin_j2000_g - deltawin_j2000_i) < 0.0003 
    OR magerr_auto_g > 0.05 ));

CREATE TEMP TABLE <tmp_bitmask> AS
    (SELECT a.* FROM <tmp_cuts> AS a
    INNER JOIN <coadd_objects_molygon> b 
    ON a.coadd_objects_id = b.coadd_objects_id
    INNER JOIN <molygon> c ON b.molygon_id_g = c.id
    INNER JOIN <molygon> d ON b.molygon_id_r = d.id
    INNER JOIN <molygon> e ON b.molygon_id_i = e.id
    INNER JOIN <molygon> f ON b.molygon_id_z = f.id
    INNER JOIN <molygon> g ON b.molygon_id_y = g.id
    WHERE c.hole_bitmask != 1 
    AND d.hole_bitmask != 1 
    AND e.hole_bitmask != 1 
    AND f.hole_bitmask != 1 
    AND g.hole_bitmask != 1);
    
CREATE <tmp_object_selection> AS
    (SELECT a.coadd_objects_id,
            a.ra, 
            a.'dec', 
            a.mag_auto_g, 
            a.mag_auto_r, 
            a.mag_auto_i, 
            a.mag_auto_z, 
            a.mag_auto_y, 
            a.magerr_auto_g,
            a.magerr_auto_r,
            a.magerr_auto_i,
            a.magerr_auto_z,
            a.magerr_auto_y 
     FROM tmp_bitmask);
    
\end{lstlisting}

7. Star-galaxy separation join:

\begin{lstlisting}
CREATE TEMP TABLE <tmp_sg_separation> AS (
    SELECT a.*
    FROM <tmp_object_selection> a
    INNER JOIN <sg_separation> b ON a.coadd_objects_id = 
    b.coadd_objects_id
    WHERE b.i='0');
\end{lstlisting}

8. Photometric redshift join:
\begin{lstlisting}
CREATE TEMP TABLE <cataog> AS (
    SELECT a.*,
           b.z_best,
           b.err_z,
    FROM <tmp_sg_separation> a
    INNER JOIN <photoz_compute> b ON a.coadd_objects_id =
    b.coadd_objects_id
    WHERE b.z_best > 0 AND b.z_best < 2.0);
\end{lstlisting}

    \begin{table*}[h!]
    \footnotesize
    \caption{Default configuration of the LSS, Cluster, GE and GA pipelines for the \textit{region selection} step.}  
    \label{tab:region_selection}
    \begin{center}
    \renewcommand{\arraystretch}{1.25}    
    \begin{tabular}{lcccc}
\hline
\hline
\textbf{Region selection parameters} & \textbf{LSS} & \textbf{Cluster} & \textbf{GE} & \textbf{GA}$^{\dagger}$ \\
%\hline
HEALpix map resolution (\texttt{NSIDE}) & 4096 & 4096 & 4096 & 4096\\
\hline
\textbf{\texttt{detfrac} maps } &  &  &  &  \\
%\hline
\texttt{detfrac} in g & None & None & None & None\\
\texttt{detfrac} in r & None & None & None & None \\
\texttt{detfrac} in i & 0.8  & 0.8 & 0.8 & 0.8 \\
\texttt{detfrac} in z & None & None & None & None\\
\texttt{detfrac} in Y & None & None & None & None\\
\hline
\textbf{Bad Region mask} &  &  &  &  \\
%\hline
1 - High density of astrometric discrepancies & No & No & No & No \\
2 - 2MASS moderate star regions ($8<J<12$) & Yes & Yes & Yes & No \\
4 - RC3 large galaxy region ($10<B<16$) & Yes & Yes & Yes & Yes\\
8 - 2MASS bright star regions ($5<J<8$) & Yes & Yes & Yes & Yes\\
16 - Regions near the LMC & No & No & No & No\\
32 - Yale bright star regions & Yes & Yes & Yes & Yes \\
64 - High density of unphysical colors & No & No & No & No \\
128 - Globular cluster regions from \citet{Har10} catalog & Yes & Yes & Yes & Yes \\
\hline
\textbf{Survey Depth maps} &  &  &  &  \\
%\hline
Apply depth map? $^{\dagger \dagger}$ & Yes & Yes & Yes & Yes  \\
Depth map type & AUTO & AUTO & AUTO & APER4 \\ 
S/N ratio & 10-sigma & 10-sigma & 10-sigma & 10-sigma \\ 
\hline 
\textbf{Systematic maps} &  &  &  &  \\
%\hline
Minimum exptime g (s) & 90 & 90 & 90 & 90\\
Minimum exptime r (s) & 90 & 90 & 90 & 90\\
Minimum exptime i (s) & 90 & 90 & 90 & 90\\
Mininum exptime z (s) & 90 & 90 & 90 & 90\\
Minimum exptime Y (s) & None & None & None & None\\
\hline 
\textbf{Additional masking} &  &  &  &  \\
%\hline
Radial query (list of ra, dec, radius values) & None & None & None & None \\
\hline \\
    \multicolumn{5}{l}{{$\dagger$}
    {The default GA configuration corresponds to the catalog used by the \textit{MWfitting} pipeline (see Section~\ref{sec:other_lightweight_catalogs}).}} \\
    
    \multicolumn{5}{l}{{$\dagger\dagger$}
    {The depth map applied is consistent with magnitude cuts in the \textit{object selection} step.}} \\    
    \end{tabular}
    \end{center}
    \end{table*}
    \begin{table*}[htb!]
    \footnotesize
    \caption{Default configuration parameters of the LSS, Cluster, GE and GA pipelines for the \textit{object selection} step. }
    \label{tab:object_selection}
    \begin{center}
    \renewcommand{\arraystretch}{1.25}
    \begin{tabular}{lcccc}
\hline
\hline 
\textbf{Object selection parameters} & \textbf{LSS} & \textbf{Cluster} & \textbf{GA} & \textbf{GE} \\
%\hline
Magnitude Type  (AUTO, DETMODEL, APER4, WAVG\_MAG\_PSF) & AUTO & AUTO & AUTO & WAVG\_MAG\_PSF \\
\hline 
\textbf{Magnitude cuts} &  &  &  &  \\
%\hline
Magnitude cut in g & None & None & None & 17<g<23 \\
Magnitude cut in r & None & None & None & 17<r<21 \\
Magnitude cut in i & 17.5<i<22 & 15<i<22 & 17.5<i<22 & None \\
Magnitude cut in z & None & None & None & None \\
Magnitude cut in Y & None & None & None & None \\
\hline 
\textbf{Signal-to-noise cuts} &  &  &  &  \\
%\hline
S/N cut in g & None & None & None & None \\
S/N cut in r & None & None & None & None \\
S/N cut in i & None & None & None & None \\
S/N cut in z & None & None & None & None \\
S/N cut in Y & None & None & None & None \\
\hline 
\textbf{Color cuts} &  &  &  &  \\
%\hline
g-r & -1.0<g-r<3.0 & -2.0<g-r<4.0 & -5.0<g-r<5.0 & 0<g-r<2.0\\
r-i & -1.0<r-i<2.5 & -2.0<r-i<4.0 & -5.0<r-i<5.0 & -5.0<r-i<5.0\\
i-z & -1.0<i-z<2.0 & -2.0<i-z<4.0 & -5.0<i-z<5.0 & -5.0<i-z<5.0\\
z-Y & -5.0<z-Y<5.0 & -2.0<z-Y<4.0 & -5.0<z-Y<5.0 & -5.0<z-Y<5.0\\
\hline 
\textbf{\texttt{Mangle} mask} &  &  &  &  \\
%\hline
Reference filter(s) (g, r, i, z, Y, All) & i & i & i & i  \\
\hline 
\textbf{\texttt{SExtractor} quality flags} &  &  &  &  \\
%\hline
Reference filter(s) (g, r, i, z, Y, All) & i & i & i & i \\
0 - Clean object & Yes & Yes & Yes & Yes \\
1 - The object has neighbors, bright and close enough\\ to significantly bias the AUTO photometry, or bad pixels & Yes & Yes & Yes & Yes  \\
2 - The object was originally blended with another one & Yes & Yes & Yes & Yes \\
4 - At least one pixel of the object is saturated  & No & No & No &  No\\
8 - The object is truncated (too close to an image boundary) & No & No & No & No\\
16 - Object aperture data are incomplete or corrupted & No & No & No &  No\\
32 - Object isophotal data are incomplete or corrupted & No & No & No &  No \\
64 - A memory overflow occurred during deblending & No & No & No & No \\
128 - A memory overflow occurred during extraction & No & No & No & No \\
\hline 
\textbf{Additional cuts} &  &  &  &  \\
%\hline
Remove artifacts associated with stars close to saturation & Yes & Yes & Yes & Yes\\
Remove objects with bad astrometric colors & Yes & Yes & Yes & Yes\\
Select objects that were observed at least once in griz & Yes & Yes & Yes & Yes\\
Remove objects in which the spreadmodel fit failed & Yes & Yes & Yes & Yes \\
\hline 
\textbf{Star--galaxy separation} &  &  &  &  \\
%\hline
Method$^{\dagger}$ & Y1 \texttt{MODEST} v2 & Y1 \texttt{MODEST} v2 & Y1 \texttt{MODEST} v2 & Y1 \texttt{MODEST} v2 \\
Reference filter(s)  (g, r, i, z, Y, All) & i & i & i & i \\
\hline 
\textbf{Photometric redshift} &  &  &  &  \\
%\hline
Method$^{\dagger}$ & MLZ/TPZ & MLZ/TPZ & MLZ/TPZ & None \\
zmin & 0 & 0 & 0 & - \\
zmax & 2.0 & 2.0 & 2.0 & - \\ 
\hline \\
    \multicolumn{5}{l}{{$\dagger$}
    {The methods for star-galaxy separation and photo-$z$ are set by selecting the corresponding input data products.}} \\  
    \end{tabular}
    \end{center}
    \end{table*}
    \begin{table*}[h!]
    \footnotesize 
    \caption{System default columns for the LSS, Cluster, GE and GA pipelines. }  
    \label{tab:column_selection}
    \begin{center}
    \renewcommand{\arraystretch}{1.25}    
    \begin{tabular}{cccc}
    \hline
    \hline
    \textbf{LSS} & \textbf{Cluster} & \textbf{GE} & \textbf{GA} \\
    \hline
       \texttt{COADD\_OBJECTS\_ID} & \texttt{COADD\_OBJECTS\_ID} & \texttt{COADD\_OBJECTS\_ID} & \texttt{COADD\_OBJECTS\_ID}\\
     \texttt{RA} & \texttt{RA} & \texttt{RA} & \texttt{RA} \\
     \texttt{DEC} & \texttt{DEC} & \texttt{DEC} & \texttt{DEC} \\
     \texttt{MAG\_[GRIZY]} & \texttt{MAG\_[GRIZY]} & \texttt{MAG\_[GRIZY]} & L\\
     \texttt{MAGERR\_[GRIZY]} & \texttt{MAGERR\_[GRIZY]} & \texttt{MAGERR\_[GRIZY]} & B\\
     \texttt{Z\_BEST} & \texttt{Z\_BEST} & \texttt{Z\_BEST} &  \texttt{MAG\_[GRIZY]} \\
     \texttt{ERR\_Z} & \texttt{ERR\_Z} & \texttt{ERR\_Z} & \texttt{MAGERR\_[GRIZY]} \\
      & & \texttt{MAG\_ABS\_[GRIZY]} & \\ 
      & & \texttt{K\_COR\_[GRIZY]} & \\ 
      & & \texttt{MASS\_BEST\_[GRIZY]} & \\ 
      & & \texttt{SFR\_BEST\_[GRIZY]} & \\ 
      & & \texttt{SSFR\_BEST\_[GRIZY]} & \\ 
      & & \texttt{AGE\_BEST\_[GRIZY]} & \\ 
      & & \texttt{EBV\_BEST\_[GRIZY]} & \\ 
    \hline \\
    \end{tabular}
    \end{center}
    \end{table*}
%
 
%%%%%%%%%%%%%%%
     \section{User Interfaces}
    \label{app:user_interfaces}
%%%%%%%%%%%%%%%

\begin{figure*}
    \centering
    \includegraphics[width=0.6\textwidth]{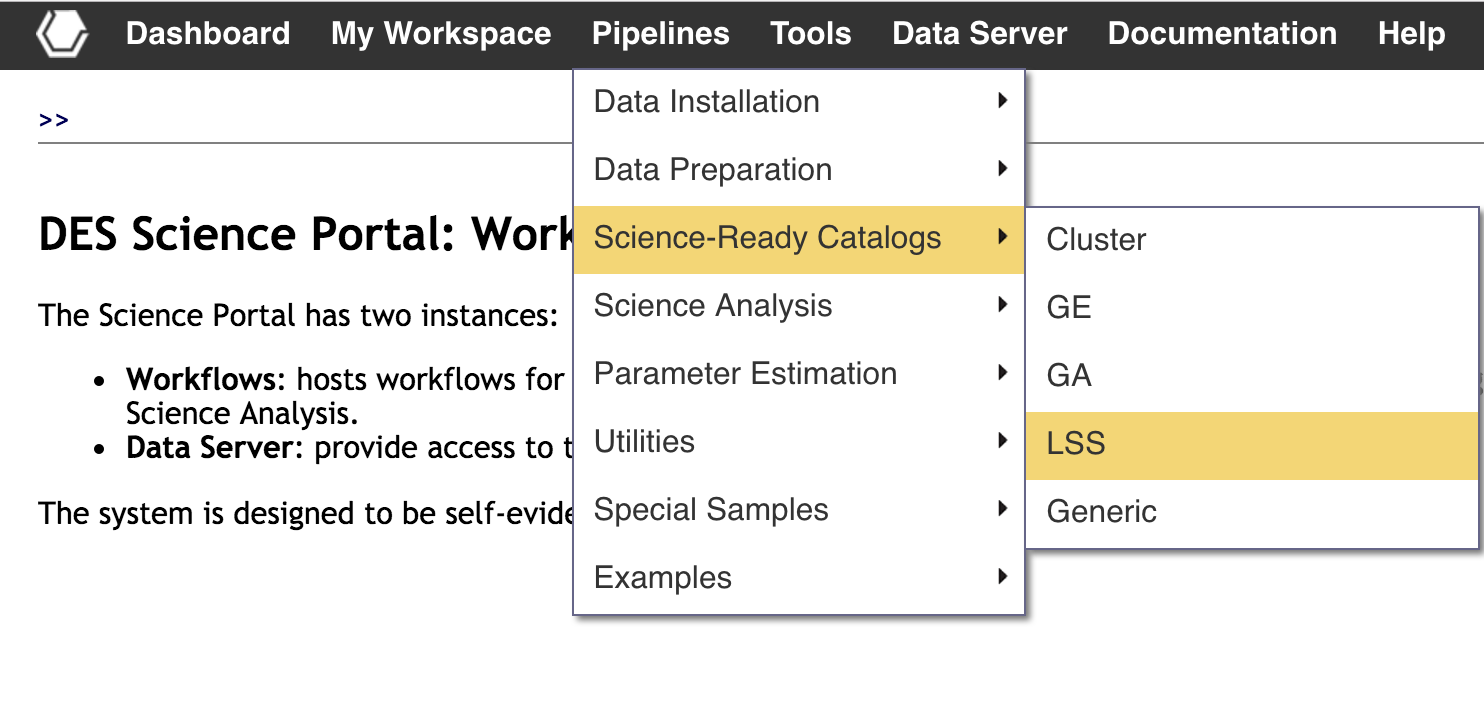}
    \caption{User interface showing the science-ready catalogs menu and the different pipelines available.}
    \label{fig:menu}
    \end{figure*}

Here we describe the user interfaces for creating science-ready catalogs as currently implemented in the portal. \footnote{See a video illustrating the creation of a science-ready catalog at \url{https://youtu.be/cZYZ6Ht0cGM}}. Figure~\ref{fig:menu} shows the science-ready catalogs menu with the different pipelines:  LSS, Cluster, GE, GA, and Generic. Once a pipeline is selected, a wizard interface leads the user through the input data selection, catalog configuration, and process submission steps. 

\begin{figure*}
    \centering
    \includegraphics[width=0.9\textwidth]{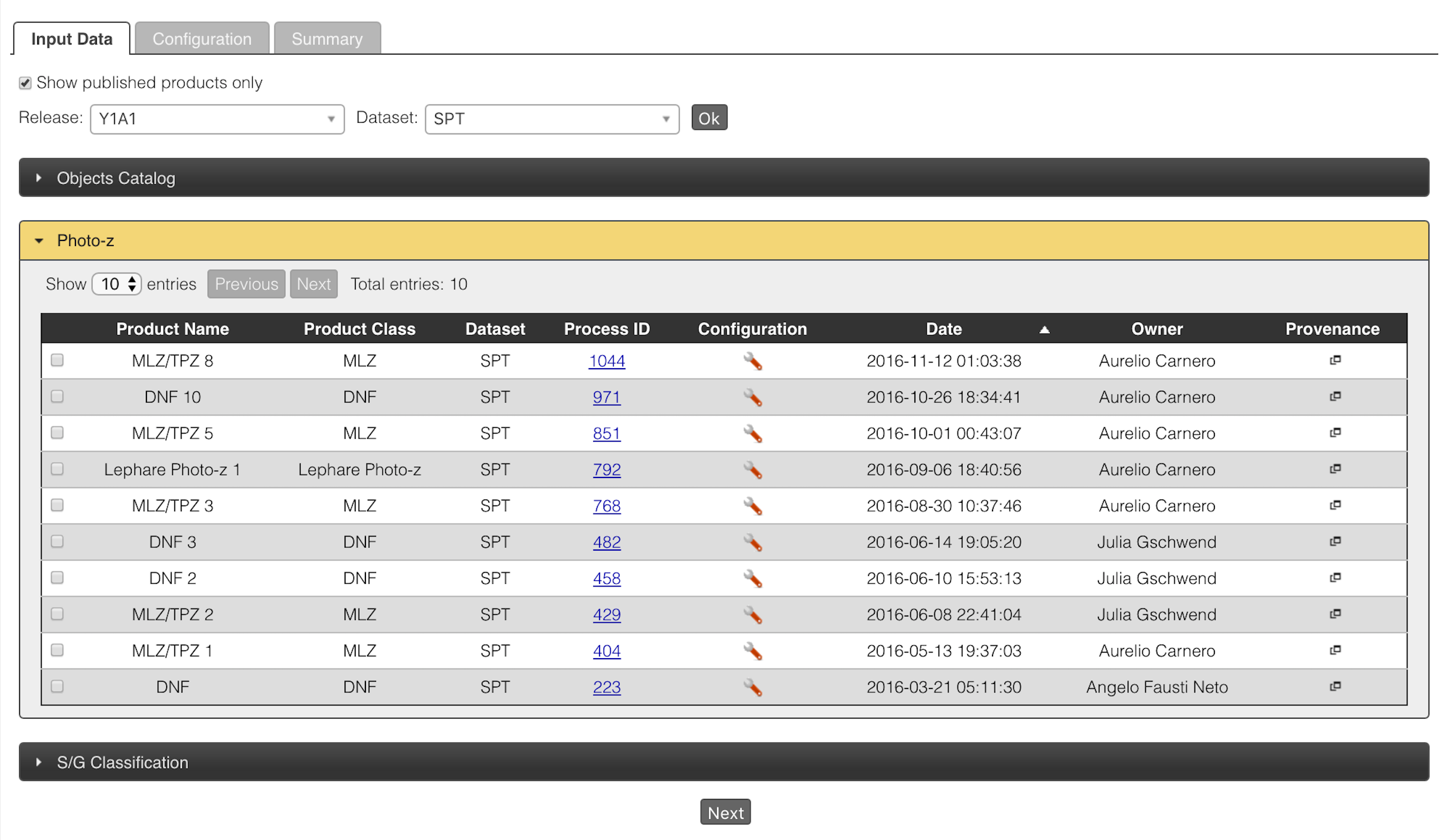}
    \caption{User interface for selecting the input data showing the available photo-$z$s, one of the value-added data products used to create  science-ready catalogs.}
    \label{fig:input_data}
    \end{figure*}

As described in Section~\ref{sec:input_data_produts}, there are several data products involved in the creation of a science-ready catalog. In addition, we want to support multiple data releases and multiple versions of a given data product with various configurations. This approach sometimes results in hundreds of options. Figure~\ref{fig:input_data} shows the user interface designed to simplify the input data selection. The user starts by filtering the available data products by Release and Dataset. In the interface, the result set is grouped by type, in this case, Objects Catalog, Photo-z and Star-galaxy separation. Once the Release and Dataset are selected, there are still many input data options for each product type. For instance, the figure lists the available Photo-$z$s identified by class, related to the Photo-$z$ method used. In order to help further the input data selection, the Process ID, Configuration, Creation Date, Owner and Provenance information are available for each option. Finally, the product types available for selection are pipeline dependent, so data products that are fixed by Data Release and Data Set, (such as the ancillary maps), are internally discovered by the {\verb query_builder }, simplifying the data selection done by the user.

\begin{figure*}
    \centering
    \includegraphics[width=0.9\textwidth]{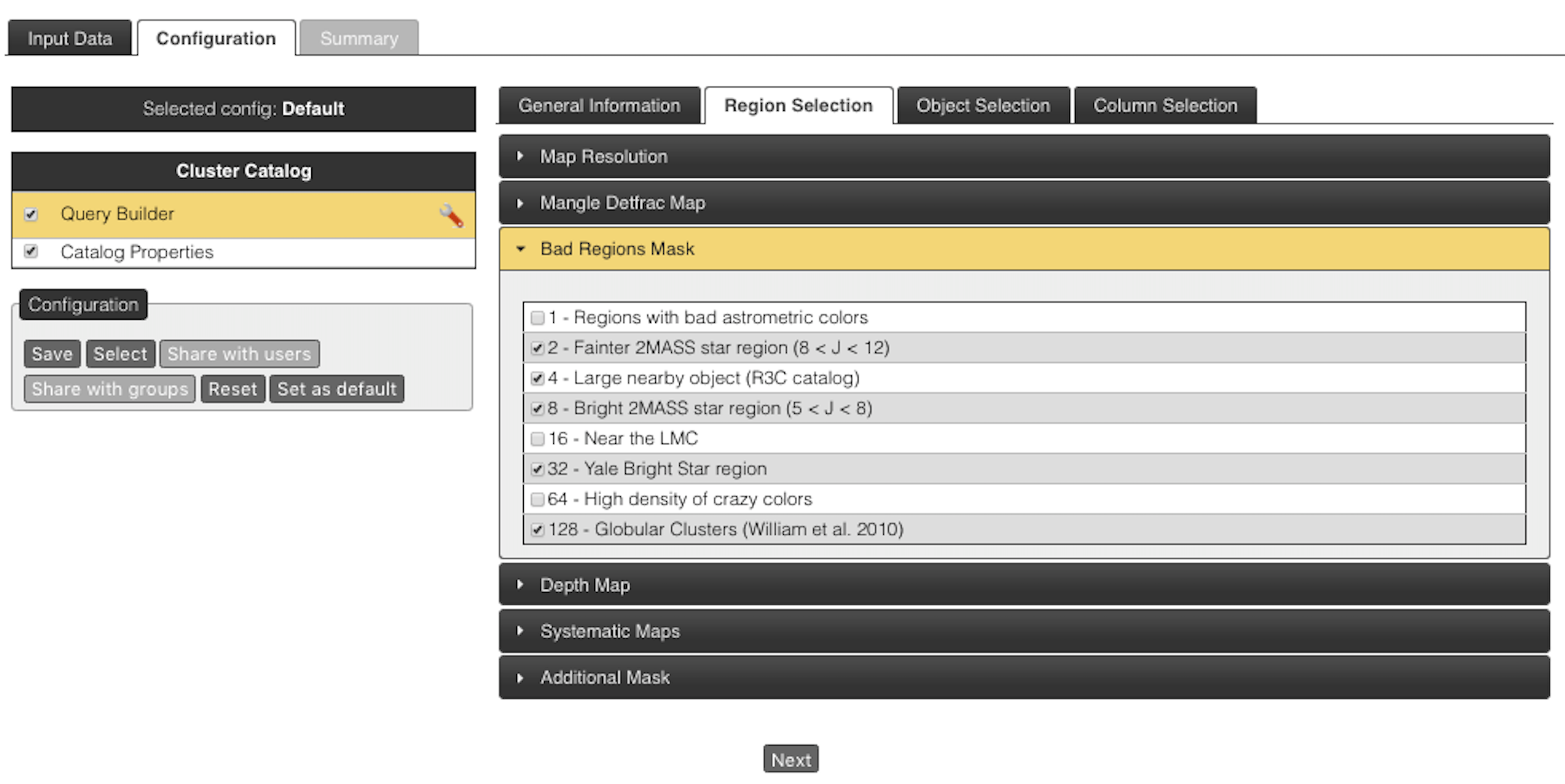}
    \caption{Configuration manager interface showing the configuration of the \textit{region selection}, one of the steps executed during the creation of a science-ready catalog.}
    \label{fig:configuration}
    \end{figure*}

The next step is the catalog configuration. Figure~\ref{fig:configuration} shows the configuration interface with General Information about the catalog being created and the configuration parameters for the \textit{region selection}, \textit{object selection} and \textit{column selection} steps. Configurations can be saved, loaded and set as default. There is also a system default configuration that can be restored using the reset button in the configuration manager interface. As a starting point, either the configuration that was set as default (if any) or the system default configuration is presented to the user (see ~\ref{app:query_builder}). Currently, there are about fifty configuration parameters for the \texttt{query\_builder}, showing the importance of having the configuration manager to keep multiple configurations for each pipeline. After the configuration step, a summary table showing the selected options is presented to the user along with a button to submit the process to create the science-ready catalog.

Given the present infrastructure, it is expected that several catalogs with different input data selections and configurations will be created making it very hard to keep track of all of them without a proper tool. That motivated the development of a dashboard, which has been successfully used to monitor the processes and the data products created in the different stages of the portal. In particular, for the \textsc{Science-ready Catalog} stage, it is possible to list all the processes that created catalogs for a given Release, Dataset and pipeline, and from that list access information about the process execution. Examples of accessible information are: process start time and duration, owner, status, if it was saved, shared or published, provenance of the data products used as input, comments made by different users on each process, a detailed process log, the products created by the process and it is also possible to export the products to other instances of the portal. In the current operation model, the export tool is the mechanism used to export the catalogs created at LIneA to the DES Science Database at NCSA.

    \begin{figure*}
    \centering
    \includegraphics[width=0.9\textwidth]{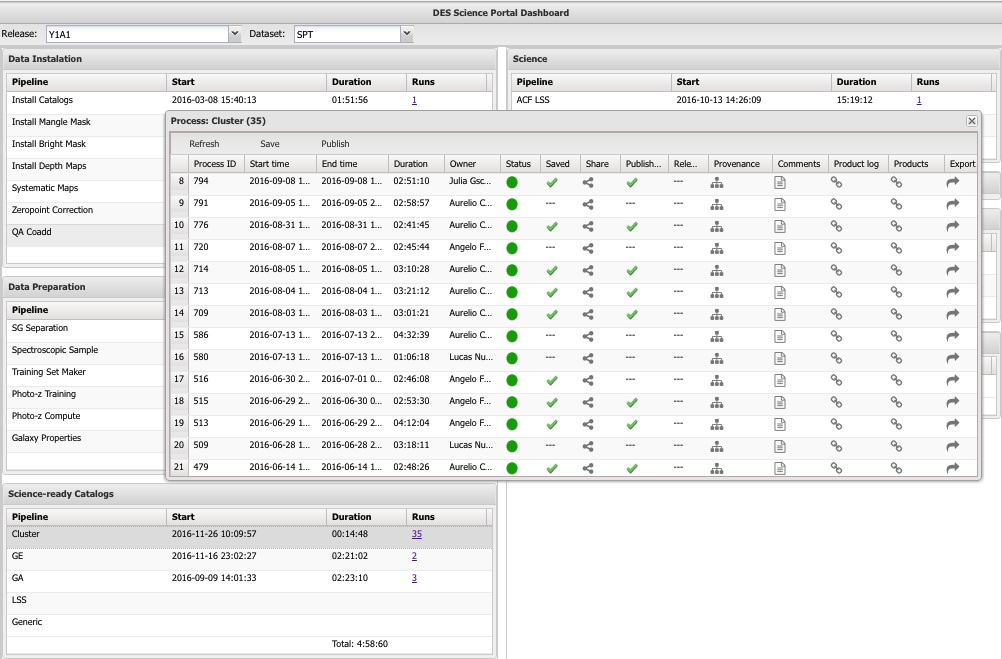}
    \caption{Dashboard interface showing the pipelines implemented in the portal and their stages. The pop-up lists all the processes that created Cluster catalogs for the Y1A1 Data Release and SPT Dataset and the associated information available for each process.}
    \label{fig:dashboard}
    \end{figure*}
    \section*{References}

%\bibliography{mybibfile}
%\bibliographystyle{model2-names.bst}
\bibliography{main}

\label{lastpage}
\end{document}